**Funding, authorship patterns and citation impact of articles funded by Ukrainian agencies before and during Russia's full-scale war (2020-2023)**

Myroslava Hladchenko, Centre for R&D Monitoring, Faculty of Social Sciences, University of Antwerp, Belgium
hladchenkom@gmail.com, Myroslava.Hladchenko@uantwerpen.be

Rodrigo Costas, CWTS, Leiden University, the Netherlands

## Abstract
This study explores funding, authorship patterns and citation impact of articles funded by the Ministry of Education and Science of Ukraine (MESU), the National Academy of Sciences of Ukraine (NASU) and the National Research Foundation of Ukraine (NRFU). The analysis focuses on articles published in Scopus-indexed journals between 2020 and 2023. The study findings revealed that the share of articles funded by these agencies increased from 8.6% in 2020–2021 to 11.9% in 2022–2023. Foreign co-funding and international co-authorship and co-affiliations were consistently associated with higher citation impact. Foreign co-affiliations had a pronounced positive effect for MESU-funded articles in 2022–2023, where their FNCI exceeded that of articles jointly funded by MESU and foreign agencies. NASU funding is associated with only modest differences in citation impact relative to unfunded articles. These effects are small and not consistently significant across authorship patterns, and they weaken in relative terms in 2022–2023 as unfunded articles partially converge in citation performance. While the results should be interpreted as average group-level tendencies rather than deterministic effects, they raise important questions about the effectiveness of current funding allocation mechanisms and evaluation criteria, highlighting the need for evidence-based reform of Ukraine's research funding system.
**Keywords:** Ukraine, research project funding, Ministry of Education and Science of Ukraine, National Academy of Sciences of Ukraine, NRFU

## 1. Introduction
Research funding agencies contribute to advancing science in general, and national science systems in particular (Möller et al., 2016). They provide additional resources beyond the block funding to universities and public research organisations. Project funding can be viewed as a governance instrument for achieving national science goals. Publications reporting funding support usually contribute to a general increase in national research performance (Möller et al., 2016), and the citation impact of funded papers is usually higher than that of unfunded papers (Möller, 2019; Wang & Shapira, 2015; Roshani et al., 2021). The efficiency of research funding allocation is measured through the publication outcomes of research projects (Lane & Bertuzzi, 2011). Exploration of funding acknowledgements enables the monitoring of publication outcomes for research projects and provides information on how to optimise the allocation of research project funding (Álvarez-Bornstein & Bordons, 2021).

Countries differ in their funding allocation models. While block research funding is considered a German model, dominance of project research funding refers to the USA's model (Capano, 2011). As a rule, the national research funding systems combine both approaches (Wand et al., 2018). Since the 1980s, Europe has increasingly shifted toward project-based research funding, a trend closely linked to the implementation of the New Public Management model (Auranen & Nieminen, 2010).

Efficient research funding allocation is important for any country, but it is even more crucial for an invaded country like Ukraine. Ukraine is a former Soviet country, which transition from the Soviet period was marked by the capture of the governmental institutions by powerful actors from the Soviet period as well as actors from the shadow economy (Riabchuk, 2009; Kudeila, 2012). With the oligarchy dominating both the economy and politics, state agencies engaged in exploitative rent-seeking. In 2013, Ukrainians started a Revolution of Dignity because the pro-Russian president refused to sign an Association Agreement with the EU. Ukrainians rebelled against the exploitation of governmental institutions, which implied a high level of corruption and disregard of law (Umland et al., 2014). Ukrainians won the Revolution of Dignity. However, the unwillingness of Russia to lose control over Ukraine resulted in Russia's invasion of Ukraine in 2014 (Kazdobina et al., 2024; Kuzio, 2022; Oleksiyenko et al., 2021). In 2022, the scope of Russia's invasion extended to all the territory of Ukraine. Since 2022, everyday survival has taken priority over research in the lives of Ukrainian scholars (Oleksiyenko, 2024). 1,518 Scholars have joined the army, and 70 of them were killed (Cabinet of Ministers of Ukraine, 2024). Between 2021 and 2022, the number of scholars in Ukraine decreased by 22% from 68,500 to 53,200 (State Statistics Service of Ukraine, 2024). Those who are not in the army, as well as other Ukrainians, experience regular shelling and power blackouts, which implies an absence of electricity, mobile and internet connections (CNN, 2024; UNESCO, 2024). The war has severely impacted Ukraine's economy, with a significant portion of the state budget redirected to the defence sector. In 2022, Ukraine's GDP declined by 29.1%, while inflation reached 26% (BBC, 2024; OECD, 2022).

Under these conditions, the efficient allocation of research funding is crucial for enabling Ukrainian science to recover and remain resilient during the war. In turn, science is a fundamental pillar of a sustainable economy, enabling the preservation and development of human capital (Scientific Committee of the National Board of Ukraine on Science and Technology, 2023). For decades, Ukrainian science system suffered from underfunding, and inefficient research funding allocation (Scientific Committee of the National Board of Ukraine on Science, Development and Innovations, 2023**;** Higher Education in Ukraine, 2019). Both national (Scientific Committee of the National Board of Ukraine on Science and Technology, 2023; De Rassenfosse, Murovana, & Uhlbach, 2023; Bezvershenko & Kolezhuk, 2022; Oleksiyenko, 2022) and international experts (Horizon 2020 Policy Support Facilities, 2017; Schiermeier, 2019) claim that Ukraine needs to break away from the Soviet-style structures, culture and practices which hold Ukrainian education and science back. This includes adopting more effective and transparent procedures for allocating project-based research funding to ensure and promote higher-quality research (Bezvershenko & Kolezhuk, 2022).

This study aims to explore funding, authorship patterns and citation impact of articles funded by the Ministry of Education and Science of Ukraine (MESU), the National Academy of Sciences of Ukraine (NASU) and the National Research Foundation of Ukraine (NRFU). The analysis focuses on articles published in Scopus-indexed journals between 2020 and 2023, encompassing two years before and two years during Russia's full-scale invasion of Ukraine.

## 2. Relationship between funding, authorship patterns and citation impact

Since 2008, Web of Science has included funding acknowledgements in scientific publications, enabling researchers to examine the sources of funding that support the production of new scientific (Costas & van Leeuwen, 2012; Paul-Hus et al., 2016; Paul-Hus et al., 2017). Later, Scopus also included these data. Though funding acknowledgements are a valuable source of data about the effects of funding policies on the science system, authors do not always acknowledge their funders (Costas & Yegros-Yegros, 2013). Álvarez-Bornstein et al. (2017), in their analysis of the completeness of funding acknowledgements in the Web of Science, found that 12% of the articles in their dataset did not include any funding information.

The presence of funding in publications of different countries is associated with an increase in national research performance. Möller, Schmidt and Hornbostel (2016) explored funding acknowledgements and highlighted the role of the German Excellence Initiative in the German research system.

Funding agencies are expected to support the most robust research proposals; however, they may prioritise projects that align with their own research agendas. While project-based funding offers significant benefits, researchers also need consistent block funding to maintain flexibility and take risks in exploring new ideas (Wang & Shapira, 2015).

Acknowledgement practices vary depending on the research areas (Grassano et al., 2017; Roshani et al., 2021; Álvarez-Bornstein & Bordons, 2021; Corsini & Pezzoni, 2023; Tian et al., 2024; Thelwall et al., 2023) and countries (Möller, 2019; Nazarovets, 2025). For example, El-Ouahi (2024) revealed that between 2008 and 2021, the percentage of papers including funding information from Middle East and North Africa (MENA) countries ranged from 40% to 70%. From a disciplinary perspective, the findings of Huang and Huang (2018) highlighted that life sciences had the highest percentage of funded papers, from 70 to 80%, depending also on the country, usually followed by agriculture & natural sciences.

Publications resulting from funding-sponsored research exhibit higher impacts in terms of both journal ranking and citation counts than research that is not sponsored (Wang & Shapira, 2015; Ou et al., 2024; Yang, 2024). Publications reporting funding acknowledgements have more citations than average (Trochim et al., 2008). At the national level, this trend was observed among Australian researchers (Sandstrom, 2009) and Canadian researchers (Campbell et al., 2010). Similar findings were reported by Roshani et al. (2021), who examined publications across three disciplines. Wang & Shapira (2015) argued that this can be because funding agencies select the best projects. Opposite to these studies, Hartel and Hooten (1992), who analysed publications in the Journal of the American Society for Information Science (JASIS), and Cronin and Shaw (1999), who examined articles from four information science journals, found no significant difference in citation rates between grant-sponsored and non-grant-sponsored research articles. These mixed findings may be attributed to variations in disciplinary focus and methodological approaches (Wang & Shapira, 2015). Research supported by multiple funding sources may undergo more rigorous review processes, potentially leading to higher quality and greater publication impact (Wang & Shapira, 2015; Rigby, 2011)

The relationship between co-authorship (the co-occurrence of more than one author in publications) and co-funding (the co-occurrence of more than one funder in one publication) is found to be positive. Álvarez-Bornstein and Bordons (2021) revealed that funding agencies increase the collaboration of researchers. Research funded from multiple countries is published in more impactful journals and has a higher publication impact (Wang & Shapira, 2015).

Apart from research project funding, the scholars can gain access to additional resources (funding, research infrastructure) and networks through multiple affiliations and collaborations that lead to co-authorship (Lander, 2015; Hottenrott & Lawson, 2017; Hottenrott et al., 2021; Kwiek, 2021). Multiple affiliations are even more beneficial than co-authorship (Lander, 2015). Co-authorship allows overcoming the lack of research infrastructure, while multiple affiliations are a way to get access to the lacking research infrastructure. Multiple affiliations are common among scholars who experience a lack of institutional support and resource constraints, which forces them to seek resources and funding from other institutions (Hottenrott & Lawson, 2017). The negative ethical implication of multiple affiliations is that they distort the measurement of institutional performance (Hottenrott & Lawson, 2017; Safaei et al., 2016; Gingras, 2014).

## 3. Context of Ukraine

### 3.1 Science system in Ukraine

Ukraine inherited from the Soviet era a structural division between the research institutes of the National Academy of Sciences of Ukraine (NASU) and primarily teaching-oriented higher education institutions. Moreover, after 1991, the sectoral academies of sciences were established, e.g. agrarian, medical, law, pedagogical (President of Ukraine, 1992; 1993).

The pre-existing hierarchical governance structure of the academy has been preserved unchanged since the Soviet period, as well as Soviet-style administrators who had long ties with communist party leadership (Josephon & Egorov, 1997). The Board of Academicians (Presidium) runs the academy**.** The Presidium consists of 34 academicians. Funding for the Presidium is allocated through the state budget of Ukraine. In 1991, 47,000 researchers were affiliated with the NASU, while in 2023 this number decreased to 13,883 (NASU, 2023). NASU consists of research institutes that are assigned to fourteen disciplinary sections. NASU has a strong school in physics established by internationally renowned theoretical physicists: the Nobel laureate L. Landau, A. Akhiezer and L. Lifshits. Physics and engineering dominate at NASU over other disciplines in the number of researchers and research institutes (Josephon & Egorov, 1997).

Regarding research at higher education institutions, in the early 1990s, the government reclassified most Ukrainian teaching-oriented institutions as universities, thereby granting them the authority to conduct research (Parliament of Ukraine, 2002; Parliament of Ukraine, 2014). In 2014, the maximum teaching hours were decreased from 900 to 600 per year to increase the time that academics can allocate to research (Ministry of Education & Science of Ukraine, 2016; Hladchenko & Westerheijden, 2021).

### 3.2 Publication requirements in Ukraine

After 1991, Ukrainian research assessment policies considered publications only in Ukrainian journals. Since 2013, the Ministry of Education and Science, being driven by the wish of Ukrainian universities to take higher positions in international rankings has introduced publications in international journals including those indexed in Scopus and Web of Science in different types of publication requirements e.g. doctoral degrees (Ministry of Higher Education and Science, Youth and Sport of Ukraine, 2012), scientific titles of associate professor and professor (Ministry of Education and Science of Ukraine, 2016), research assessment of higher education institutions (Cabinet of Ministers of Ukraine, 2017) (Table 1). Bibliometric characteristics of researchers' output are among the criteria used for selecting projects funded by the Ministry of Education and Science of Ukraine (MESU). Applicants are required to report the cumulative h-index and total number of citations in Scopus and WoS (and in Google Scholar for social sciences and humanities) of the project leader and five main participants of the project, who also need to submit a publication list specifying the journal quartile in Scopus and WoS for each publication. Applicants also submit a list of articles published in journals that are not indexed in Scopus and WoS. The application form includes a section titled "Criteria for the Expected Outcomes of the Project." One of the items in this section pertains to articles published in Scopus- and WoS-indexed journals, categorised by journal quartiles (Ministry of Education and Science of Ukraine, 2023).

However, previous studies have shown that the introduction of publication requirements emphasising Scopus-indexed journals led to an increase in the number of such articles published by Ukrainian scholars. However, to some extent, the quantity was prioritied over quality and Ukrainian academics also extensively published in local Scopus-indexed journals (Nazarovets, 2020; 2022; Hladchenko, 2022).

**Table 1.** Summary of requirements for (international) publications in Ukrainian research assessment policies

| Policy | Publication requirements |
|---|---|
| Doctoral degrees (2013) | Publications in Ukrainian Scopus- and WoS-indexed journals or in international journals<br>Since 2020 publications in any Scopus- and WoS-indexed journals<br>Since September 2021 for the Doctor of Sciences and since September 2020 for PhD, the number of articles needed for their obtention can be reduced by publishing in Q1-Q3 journals |
| Licensing (2015) | At least five articles published during last five years in Ukrainian journals, in journals indexed in databases including Scopus and WoS |
| Requirements for scientific titles of associate professor and professor (2016) | Article(s) in Scopus- and WoS indexed journals |
| Research assessment of higher education institutions (2017) | Articles in journals indexed in databases (the legislation does not clarify the names of databases) |
| Funding of research projects (2019) | Publications in Scopus- and WoS-indexed journals |
| Performance-based funding (2019) | Indicator of international recognition is defined based on the positions in international university rankings. |

### 3.3 Integration into the European Research Area

As an associate country, Ukraine is supposed to integrate into the European Research Area (ERA). In 2021, the Ministry of Education and Science of Ukraine adopted the "Roadmap for Ukraine's Integration into the European Research Area". However, the Cabinet of Ministers of Ukraine has neither demonstrated a commitment to implementing the road map nor allocated the necessary resources (Bezvershenko & Kolezhuk, 2022). Between 2016 and 2020, Ukraine was fully associated with Horizon 2020. In 2023, Office Horizon Europe in Ukraine was established in the National Research Foundation of Ukraine (NRFU).

### 3.4 Ukrainian funding agencies

Ukrainian government allocates the major share of available research funding to the following organisations: 1) the National Academy of Sciences of Ukraine (NASU) and sectoral academies; 2) the Ministry of Education and Science of Ukraine (MESU); and 3) the National Research Foundation of Ukraine (NRFU).

NASU receives the largest share of the state research budget. Its funding consists of two components: (1) basic (institutional) funding, and (2) competitive project-based funding. The latter is allocated through a specific budget line titled “Support for the Development of Prioritised Directions of Research”, which appears in the State Budget of Ukraine. These projects usually have a duration of one to two years. The application process is bureaucratically demanding and requires prior approval of the research proposal by the Scientific Board of the applicant’s research institute (NASU, 2024).

The Ministry of Education and Science of Ukraine (MESU) distributes research funding on a competitive basis to Ukrainian higher education institutions (Ministry of Education and Science of Ukraine, 2022). It also provides financial support to the “Academician Vernadskyi” Antarctic station, and funds state prizes and scholarships for outstanding young scientists. Additionally, MESU oversees the allocation of national co-funding for participation in international research programmes, including the EU Framework Programme Horizon 2020.

Established in 2018 through the reorganisation of the former State Fund for Fundamental Research (Cabinet of Ministers of Ukraine, 2018), the NRFU began issuing competitive research grants in 2020. It administers both state budget funding and funding from international partners, such as Germany’s DFG, Poland’s FNP, and the Netherlands’ NWO, to support bilateral or multilateral research projects involving Ukrainian scientists.

Beyond these three core agencies, the Ukrainian government also allocates research funding through several line ministries (e.g. the Ministry of Finance, the Ministry of Health), and the State Fund for Regional Development (SFRD, 2021).

### 3.5 Ukrainian research funding during Russia’s full-scale war

Russia’s invasion of Ukraine in 2014 and its full-scale escalation in 2022 led to significant shifts in government priorities, with financial resources for science redirected toward urgent defence and recovery needs. While the Law of Ukraine “On Scientific and Scientific-Technical Activity” stipulates that public expenditure on science should be no less than 1.7% of GDP, actual R&D spending relative to GDP fell from 0.75% in 2010 to 0.33% in 2023 (Iurchenko & Ponomarenko, 2025). According to the UNESCO Institute for Statistics, gross domestic expenditure on research and development shrank by 38.5% between 2021 and 2022, dropping from €1,705.5 million to €1,180.0 million (in purchasing power parity, adjusted for inflation). In 2024, UNESCO estimated that restoring Ukraine’s public research infrastructure would cost at least US$1.26 billion, with costs continuing to rise as the war persists.

Since 2021, the National Academy of Sciences of Ukraine—comprising 450 affiliated institutes—has experienced a 48% budget reduction, from €201.5 million in 2021 to €115.4 million in 2023. This significant cut has led to the suspension of numerous research programmes and the weakening of many research teams (UNESCO, 2024). Despite these cuts, NASU continued to receive funding under the national programme “Support for the Development of Prioritised Directions of Research,” receiving €11.74 million in 2020 (Parliament of Ukraine, 2020), €12.34 million in 2022, and €12.87 million in 2023 (Slovo i Dilo, 2024).

Research funding administered by the Ministry of Education and Science of Ukraine (MESU) for university-based research amounted to €3.23 million in 2021, decreased slightly to €3.09 million in 2022, and then sharply declined to €0.70 million in 2023 (Slovo i Dilo, 2024). These figures reflect the severe impact of wartime budget constraints on research in Ukrainian higher education institutions. Meanwhile, no funds were allocated in 2020 due to emergency budget cuts amid the COVID-19 pandemic (Slovo i Dilo, 2024).

The National Research Foundation of Ukraine (NRFU), established in 2018 to competitively allocate research grants, received approximately €9.1 million in 2020 (Parliament of Ukraine, 2020), increasing to €22.6 million in 2021 (Slovo i Dilo, 2024). Although a similar allocation of around €23.3 million was planned for 2022, no grants were disbursed due to the reallocation of funds for defence following the full-scale invasion in February. During this period, NRFU sought support through international fundraising campaigns. Funding was partially restored in 2023 at approximately €17.9 million, which covered only about 60% of its obligations to winning projects (Science Europe, 2023). Among three Ukrainian agencies, the NASU has received the largest share of state funding. In 2023, NASU accounted for 61.2% of state research funding, while MESU and NRFU combined received 16.9% (Ministry of Education and Science of Ukraine, 2024).

### 3.6 Supporting academic mobility schemes for Ukrainian scholars fleeing from the war

In an effort to support Ukrainian scholars, the European Commission, national governments, and universities have provided funded positions and fellowships — for example, through MSCA4Ukraine, SNSF, the Volkswagen Foundation, and the Alexander von Humboldt Foundation. The U.S. National Academies of Sciences, in partnership with the Polish Academy of Sciences, launched an initiative to help Ukrainian researchers settle in Poland (OECD, 2022). This funding opportunity — along with the long-standing collaboration between Ukrainian and Polish academics, which intensified after the Russian invasion of Ukraine in 2014 — have contributed to academic migration to Poland (Hladchenko, 2025). Geographical proximity and established academic contacts made Poland particularly attractive (Kiselyova & Ivashchenko, 2024). In the UK, several universities have participated in the "Researchers at Risk Fellowships"—organised by the British Academy and the Council for At-Risk Academics (CARA)—providing Ukrainian academics displaced by the war with fully funded placements in UK institutions for up to two years.

## 4. Data and methods

***Data***

Since 2013, the research assessment policies in Ukraine have prioritised publications in Scopus; therefore, data for this study were retrieved from the Scopus database. We collected all articles (N=61,698) published between 2020 and 2023, in which at least one author indicated an affiliation with Ukraine. By 'funded paper' we mean publications that acknowledge the support of a research sponsor or grant provider, including partial support such as provision of instruments or equipment, which is common in fields like medicine.

We acknowledge the potential limitations in identifying Ukrainian research funding due to inconsistencies and vagueness in acknowledgements provided by authors. Scopus metadata distinguish between "Funding Details" (structured information on the funding agency) and "Funding Text" (the full acknowledgement statement). In cases where "Funding Details" lacked sufficient information, we cross-checked the corresponding "Funding Text" to determine whether financial support had indeed been provided and by which agency.

Some acknowledgements included vague language that made it difficult to distinguish between expressions of institutional gratitude and actual financial support. We applied careful judgment to classify only those cases that clearly involved financial backing as funded. In several instances, authors cited only the project registration number without naming a funding agency. For such cases, we consulted the official Ukrainian Institute of Scientific and Technical Expertise and Information (UKRNTI) repository, which under Ukrainian law includes all state-funded research projects (Ministry of Education and Science of Ukraine, 2022a). This registry provides information on the funding agency and the allocated budget.

Additionally, we identified cases in the UKRNTI database where projects were marked as "initiated by academics," indicating the absence of financial support from the state or the host university. Such publications were not classified as funded. We also excluded cases where Scopus had incorrectly identified a government agency mentioned in a non-funding context (e.g., the Pension Fund of Ukraine) as a funding source.

Among the 61,698 articles published by Ukrainian scholars between 2020 and 2023, we identified 15,179 (25%) as having received funding. This study specifically focuses on articles funded by the Ministry of Education and Science of Ukraine (MESU), the National Academy of Sciences of Ukraine (NASU), and the National Research Foundation of Ukraine (NRFU).

As this study aims to explore the relationship between funding, authorship patterns and citation impact, the following funding and authorship patterns were analysed. The *funding patterns* include: a) acknowledgement to one of three Ukrainian agencies: Ministry of Education and Science of Ukraine (MESU), National Academy of Sciences of Ukraine (NASU), National Research Foundation of Ukraine (NRFU); b) acknowledgement to both one of three Ukrainian agencies and one or more foreign funding agencies (MESU&FF, NASU&FF, NRFU&FF); c) acknowledgement only to one or more foreign funding agencies (FF); d) articles which do not acknowledge any funding agency (either Ukrainian or international) (WF).

The literature indicates that both foreign co-affiliation and foreign co-authorship provide scholars with access to additional resources and research infrastructure, which is associated with advantages for the resulting publications compared to domestically authored articles (Lander, 2015). However, prior research suggests that the benefits related to foreign co-affiliation tend to outweigh those associated with foreign co-authorship, as multiple affiliations facilitate more direct and sustained access to external research infrastructure and funding (Hottenrott & Lawson, 2017; Hottenrott et al., 2021). Based on this literature, the authorship patterns analysed in this study include foreign co-affiliation (FCAF), foreign co-authorship (FCAU), and Ukrainian-only affiliation (UKR).

In this study, foreign co-affiliation refers to articles for which at least one author reports affiliation with both Ukrainian and foreign institution. Foreign co-authorship refers to articles for which at least one author reports affiliation exclusively with a foreign institution. Ukrainian affiliation denotes articles authored solely by researchers affiliated with Ukrainian institutions. The CWTS Scopus in-house database contains country- and institution-level affiliation information for each author, which enables systematic identification of these authorship patterns and consistent classification of articles into the three categories. Articles displaying both foreign co-affiliation and foreign co-authorship were counted twice being assigned to both authorship patterns, these two authorship patterns are not mutually exclusive.

*Methods*

Citations to articles in the dataset were extracted from CWTS Scopus in-house database released in March 2025. CWTS Scopus in house database is updated once a year. As articles in the dataset were published in 2020-2023, citations from 2025 provide the largest posible citation window from available data (Abramo et al., 2011). Disciplines differ in publication, citation, and authorship practices, as well as publications published earlier have citation advantages. To overcome these discripancies and ensure comparapitily of articles from different disciplines and published in different years, it is standard in bibliometrics to normalise citations (Waltman & van Eck, 2013; Bornmann & Wohlrabe, 2019). The recommendation to use normalised bibliometric indicators instead of bare citation counts is one of the ten guiding principles for research metrics listed in the Leiden manifesto (Hicks et al. 2015; Wilsdon et al. 2015).

Field-normalised citation impact (FNCI) for each article in the dataset was calculated in CWTS Scopus in-house database. Citations to each article were divided by average global number of citation to articles published in the same discipline in the same year. As such, citations to each article were normalised by discipline and year of publication. The number of citations for each article was normalised by dividing it by the global average number of citations for articles in the same discipline, publication year, based on the Scopus subject classification system. Articles classified by Scopus in more than one discipline were fractionalised among these disciplines.

We compared citation impact across various categories of funding and authorship using the mean of field-normalised citation impact (FNCI) values. Study results contain descriptive statistics of min, max, mean and median FNCI across funding agencies and authorship patterns for two period (2020-2021 and 2022-2023).

Given the overdispersion of FNCI data, a negative binomial regression model with interactions between period (2020–2021 vs. 2022–2023), fund type, and authorship category was used to explore the relationship between funding, authorship patterns, and FNCI. The intercept represents NASU-funded articles in 2020–2021 with Ukrainian authorship. The model was run in R, using the MASS package. Visualistion was done in R using ggplot2 package.

The analysis compares two publication periods—2020–2021 and 2022–2023—to assess whether observed associations among funding sources, authorship patterns, and citation impact change in magnitude or structure following the onset of the war. This does not assume that the war is the sole driver of these relationships, nor do they attempt to isolate counterfactual outcomes in the absence of war. Instead, the war period is used as a temporal marker to contextualise observed differences in research organisation and performance under dramatically altered conditions.

## 5. Results

### 4.1. Funding agencies acknowledged in articles authored by Ukrainian scholars

Fig. 1 illustrates the dynamics in the percentage of articles with various types of funding as well as those without. Between 2020 and 2023, the percentage of Ukrainian articles acknowledging funding increased from 21.7% to 28.5%. This growth was observed across all funding types Ukrainian, foreign, and combined Ukrainian and foreign with the most significant increase seen in articles supported by Ukrainian funding. It is important to note, however, that the overall research output of Ukrainian scholars declined by 3.6% in 2022–2023 compared to 2020–2021, decreasing from 31,422 to 30,276 articles.

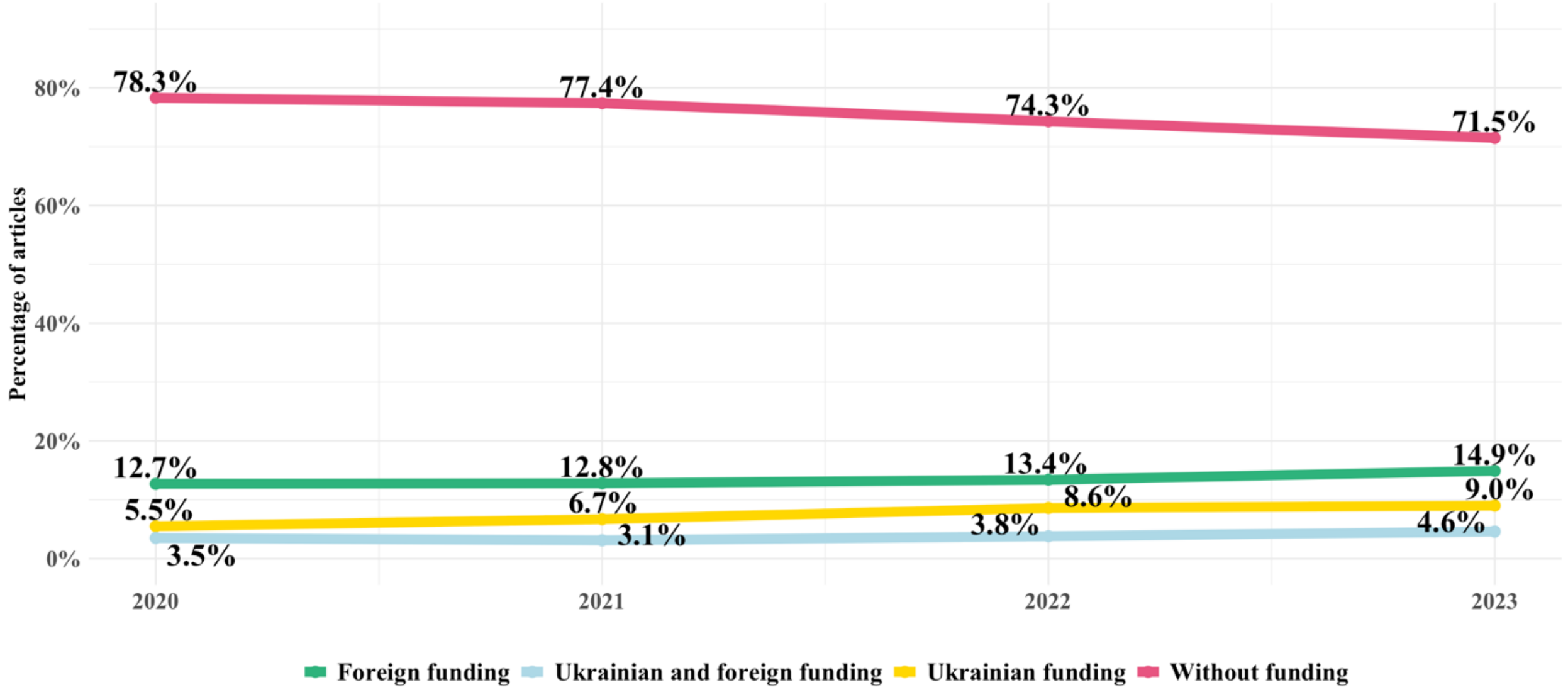

**Figure 1.** Percentage of articles with funding acknowledgements in 2020-2023.

Table 2 lists the top ten funding agencies acknowledged in Ukrainian articles published in 2020–2021 and 2022–2023. It shows that the share of articles funded by the National Academy of Sciences of Ukraine (NASU), the Ministry of Education and Science of Ukraine (MESU), and the National Research Foundation of Ukraine (NRFU) increased from 8.6% (N = 2,692) in 2020–2021 to 11.9% (N = 3,598) in 2022–2023. In 2020–2021, MESU (4.3%) and NASU (4.0%) accounted for the largest shares of funded articles, while NRFU ranked ninth with 0.9%. By 2022–2023, MESU (5.0%) and NASU (4.8%) maintained their positions as the top national funding agencies. Notably, NRFU rose from ninth to fourth place, increasing its share to 3.0%, with the number of articles acknowledging NRFU funding growing from 282 to 915.

Table 2 Top ten funding agencies in articles authored by Ukrainian scholars in 2020-2023

| **2020-2021** | | | **2022-2023** | |
|---|---|---|---|---|
| 1. | Ministry of Education and Science of Ukraine (MESU) | 4.3% (1463) | Ministry of Education and Science of Ukraine (MESU) | 5.0% (1502) |
| 2. | National Academy of Sciences of Ukraine (NASU) | 4.0% (1260) | National Academy of Sciences of Ukraine (NASU) | 4.8% (1463) |
| 3. | EU | 3.9% (1222) | EU | 4.0% (1198) |
| 4. | German Research Foundation (DFG) | 1.4% (441) | National Research Foundation of Ukraine (NRFU) | 3.0% (915) |
| 5. | Russian Foundation for Basic Research (RFBR) | 1.2% (382) | National Science Centre (NCN), Poland | 1.6% (496) |
| 6. | National Science Foundation of China (NSFC) | 1.2% (378) | German Research Foundation (DFG) | 1.6% (495) |
| 7. | National Science Centre (NCN), Poland | 1.2% (369) | National Science Foundation of China (NSFC) | 1.6% (482) |
| 8. | National Science Foundation (NSF), the USA | 1.1(334) | Federal Ministry of Education and Research (BMBF), Germany | 1.0% (315) |
| 9. | National Research Foundation of Ukraine (NRFU) | 0.9% (282) | National Science Foundation (NSF), the USA | 1.0% (312) |
| 10. | Department of Energy (DOE), the USA | 0.9% (281) | National Institute of Nuclear Physics (INFN), Italy | 0.9%(278) |

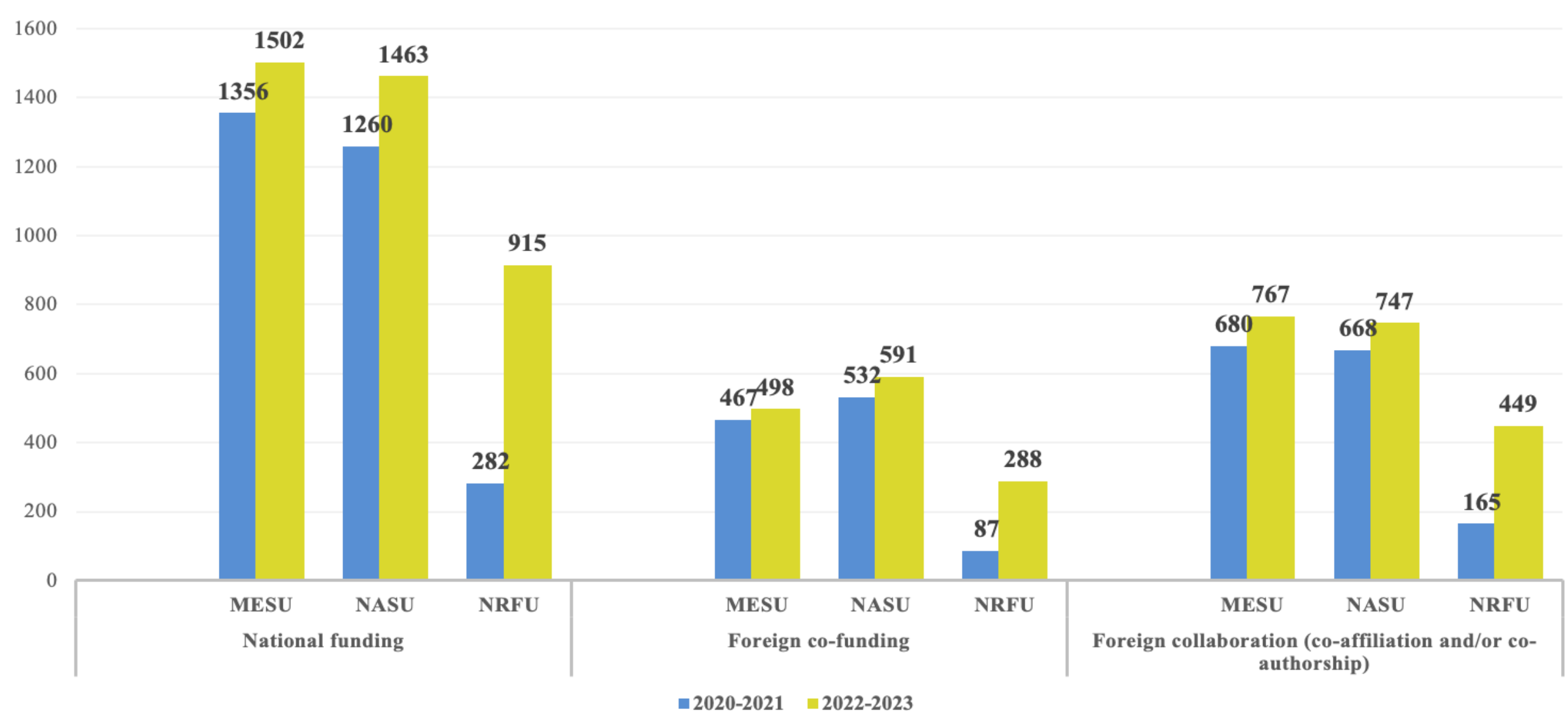

**Fig. 2.** Funding and authoship patterns in articles funded by MESU, NASU and NRFU (number).

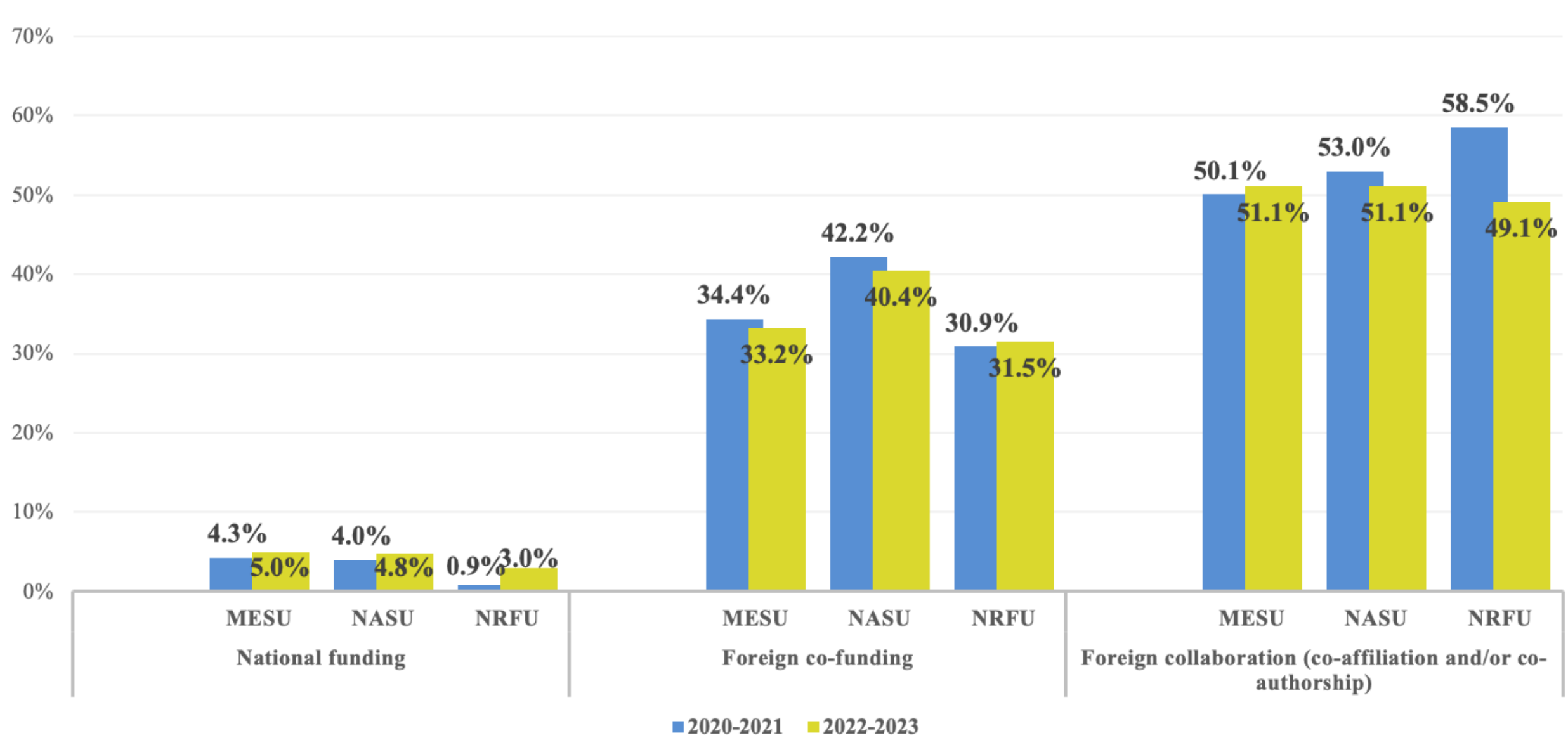

**Fig. 3.** Funding and authoship patterns in articles funded by MESU, NASU and NRFU (percentage).

Fig. 2 and 3 illustrate funding and authorship patterns in articles funded by MESU, NASU and NRFU. Fig 2 shows that in 2022-2023, the number of articles funded solely by these agencies, as well as those co-funded by foreign agencies, increased. However, Fig 3 highlights that in 2022-2023, the percentage of articles with foreign co-funding decreased slightly in MESU- and NASU-funded articles and increased in NRFU-funded articles. The highest percentage of articles with foreign co-funding (40.4%) was in NASU-funded articles. This can be atributed to physical sciences being the main discipline of the NASU, where research is often organised through large multinational consortia. The number of articles authored by Ukrainian scholars with foreign co-affiliations and internationally co-authored articles increased for all three funding agencies. However, the percentage of these articles increased in MESU-funded articles and decreased in NASU- and NRFU-funded articles. The highest percentage was observed in MESU- and NASU-funded articles (51.1%).

Figures 4 and 5 present the percentages of articles with Ukrainian affiliations, foreign co-authorship, and foreign co-affiliations across three agencies—MESU, NASU, and NRFU—for two periods: 2020–2021 and 2022–2023. Articles authored by Ukrainian scholars with foreign co-affiliations and those co-authored with foreign scholars are not mutually exclusive. The key distinction between the two figures is funding source: Figure 4 includes articles with foreign co-funding, while Figure 5 focuses on those without foreign co-funding. Figure 4 shows that foreign co-funding promotes international collaboration. Across all agencies and both periods, the share of internationally co-authored articles and those authored by scholars with foreign co-affiliations are significantly higher among articles with foreign co-funding. For instance, NASU-funded articles with foreign co-funding published in 2020–2021 show 37.7% foreign co-authorship, compared to only 12.8% without foreign co-funding (Fig. 5). Conversely, the percentage of articles with Ukrainian affiliations is much higher in the absence of foreign co-funding, with MESU having 45.5% and 45.0% in Fig. 5, compared to only 4.4% and 3.9% in Fig. 4.

In Figure 4 (with foreign co-funding), the share of articles authored by scholars with foreign co-affiliations shows a modest increase (e.g., MESU rises from 11.6% to 14.1%), while the share of articles with Ukrainian affiliation slightly decreases. Foreign co-authorship remains high but slightly declines for all agencies in 2022–2023. In Figure 5 (without foreign funding), Ukrainian affiliation remains dominant but decreases for NASU and NRFU. Notably, foreign co-affiliation increases for all three agencies. The changes observed in 2022–2023 are likely influenced by the full-scale Russian invasion, which resulted in the displacement of many Ukrainian researchers and fostered increased international collaboration between those who remained in Ukraine and scholars with foreign co-affiliations.

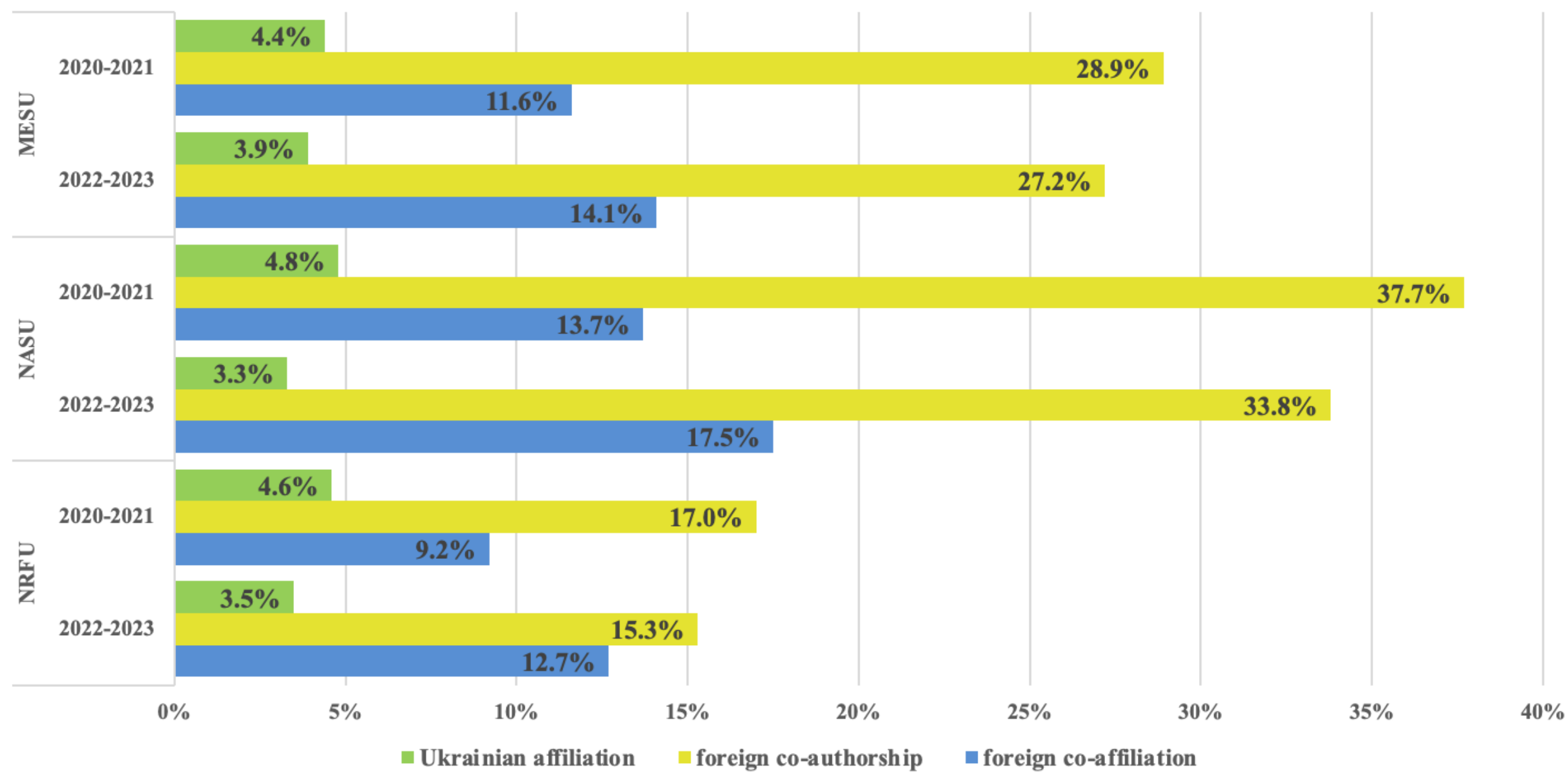

**Fig. 4.** The percentage of articles with three types of authorship in MESU-, NASU- and NRFU-funded articles with foreign co-funding of the total number of articles funded by each agency.

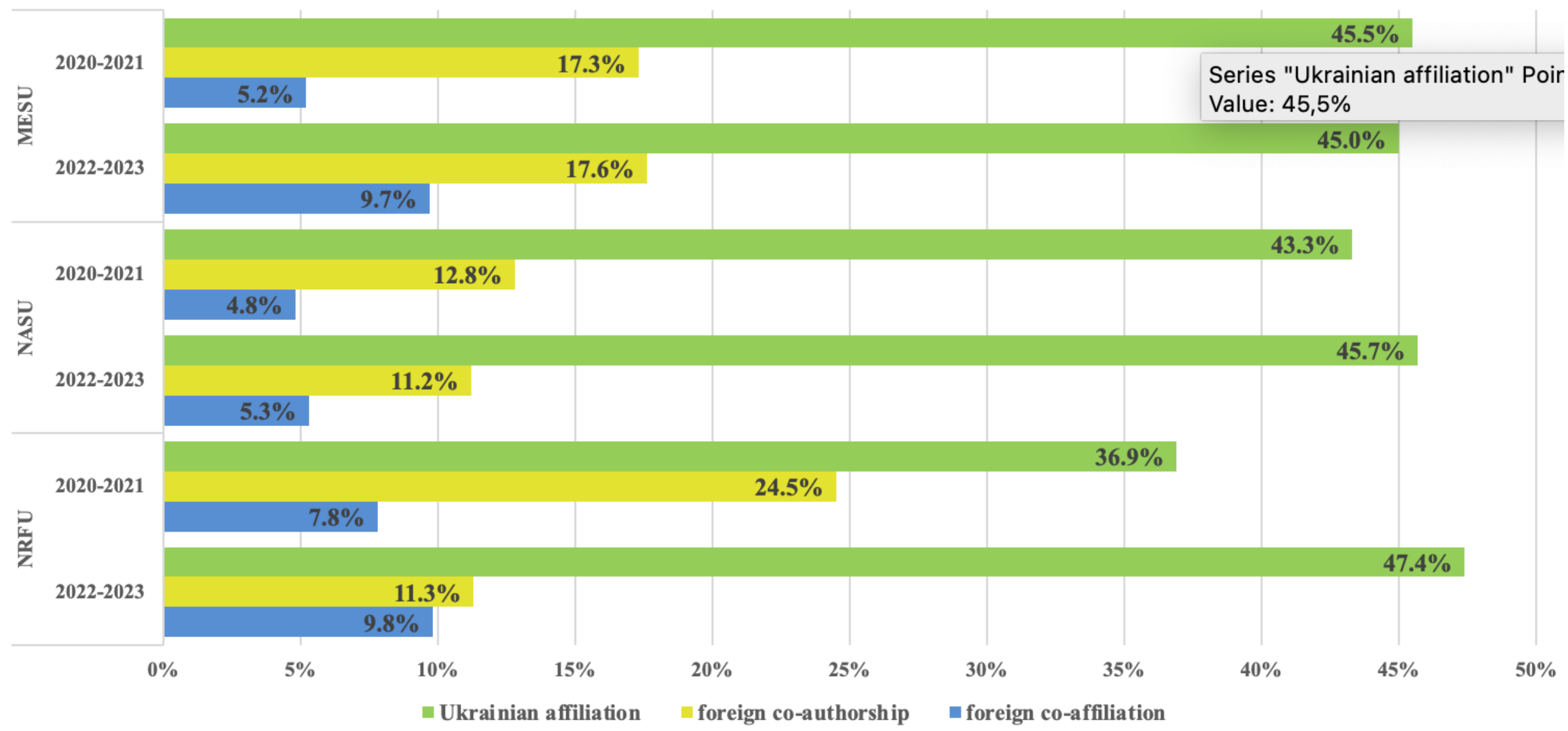

**Fig. 5.** The percentage of articles with three types of authorship in MESU-, NASU- and NRFU-funded articles without foreign co-funding of the total number of articles funded by each agency.

## 5.2. Top ten countries of foreign co-affiliation, co-authorship and co-funding in MESU-funded articles

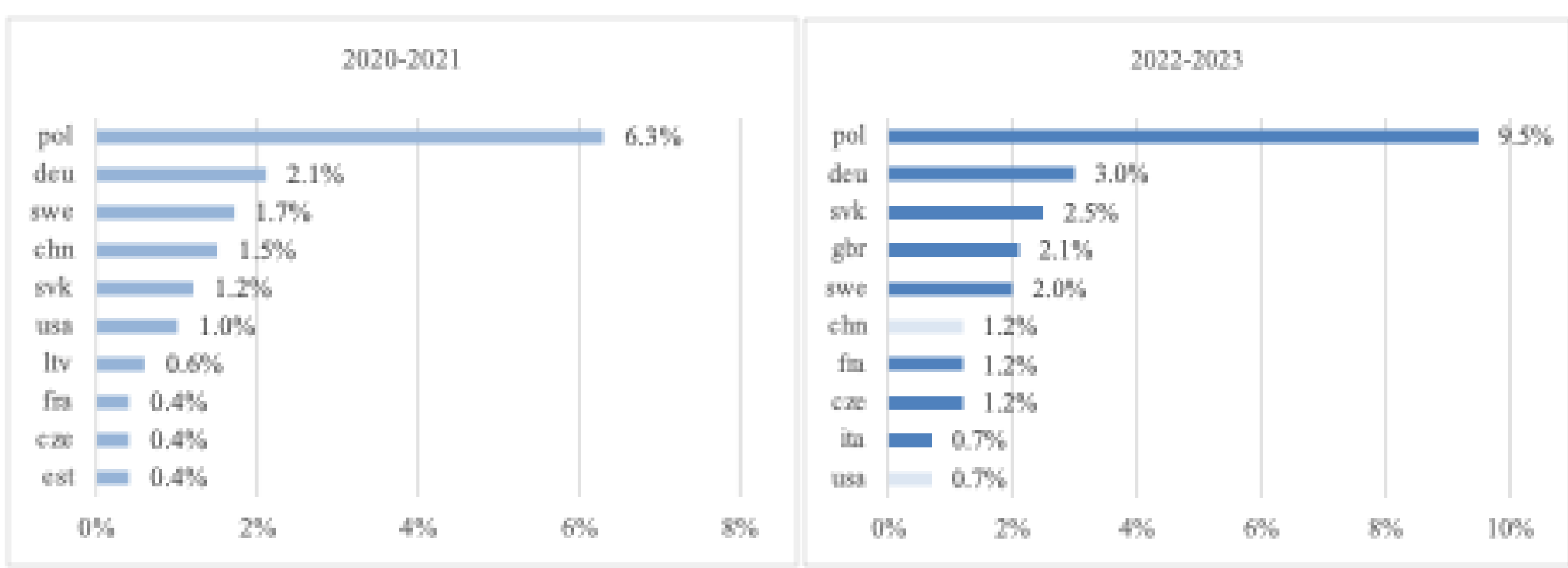

**Fig. 6.** Top ten countries of foreign co-affiliation of Ukrainian scholars in MESU-funded articles

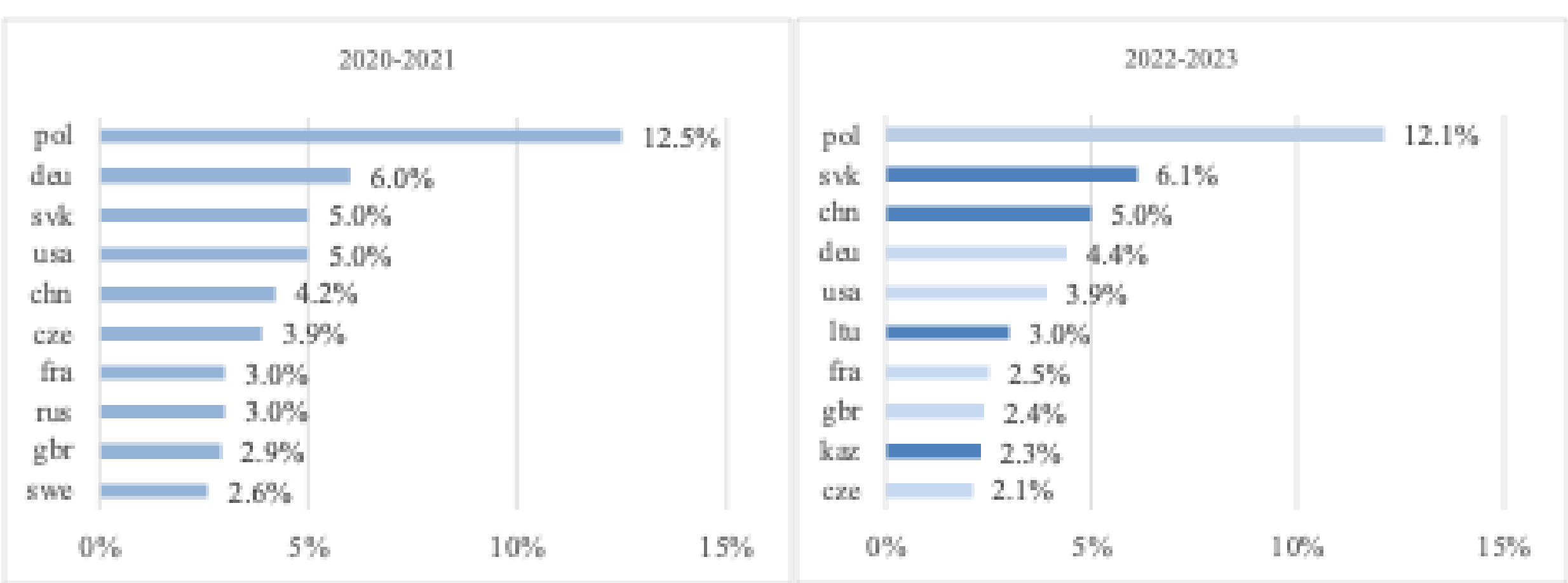

**Fig. 7.** Top ten countries of foreign co-authorship in MESU-funded articles

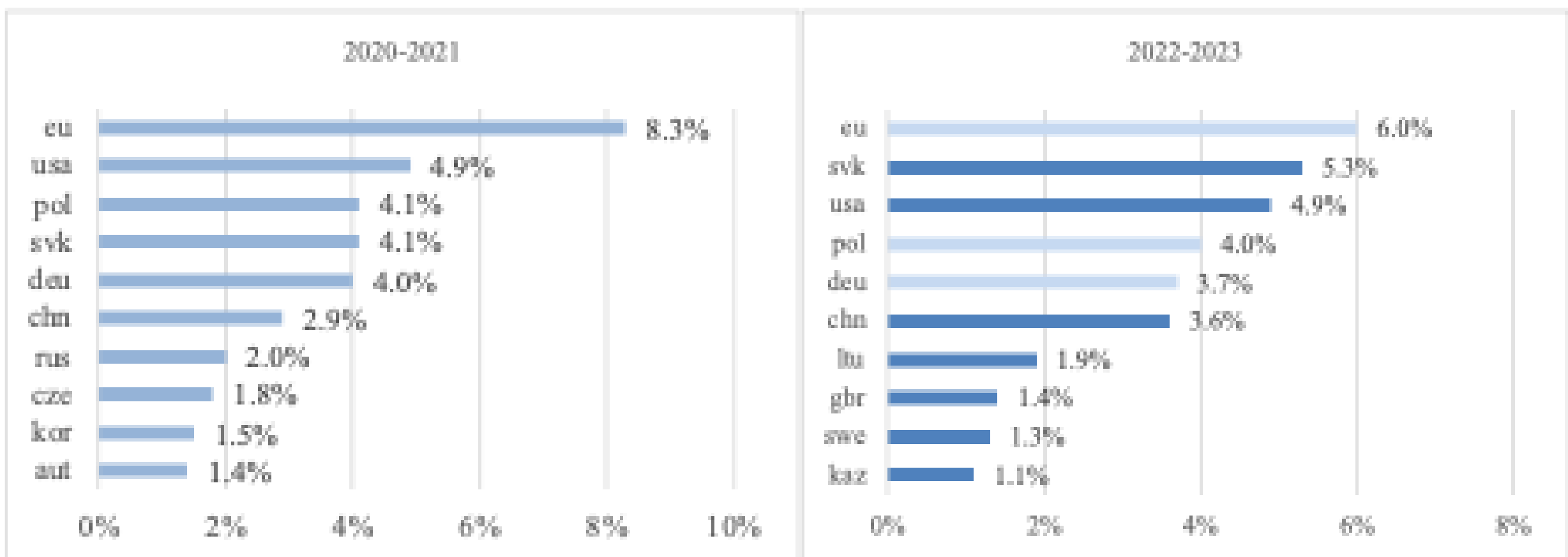

**Fig. 8**.Top ten foreign co-funding countries in articles funded by MESU

Figures 6–8 provide insights into foreign co-affiliation, co-authorship, and co-funding in MESU-funded articles during 2020–2021 and 2022–2023. Figure 6, which focuses on foreign co-affiliation, shows that Poland remained the leading country, with its share increasing from 6.3% to 9.5% in 2022–2023. This increase suggests a deepening institutional presence of Ukrainian scholars. Germany and Slovakia also experienced growth in co-affiliations, likely reflecting strengthened academic ties due to geographic proximity and targeted post-2022 support initiatives. Notably, *the UK* and *Italy* entered the list in 2022–2023, countries known to have hosted many displaced Ukrainian scholars (Deutsch-Ukrainische Akademische Gesellschaft, 2025). The continued presence of major Western countries—such as Germany, the USA, the UK, and Sweden—indicates a growing engagement of Ukrainian scholars with globally connected and well-resourced institutions, likely driven by both displacement and increased access to international support programs.

Figure 7 on foreign co-authorship reveals that Poland retained its leading position in both periods (12.5% to 12.1%). While its share slightly declined, the consistently high figure reflects sustained collaborative ties. *Slovakia, China,* and *Germany* saw stable or increasing co-authorship shares. In contrast, Russia disappeared from the 2022–2023 list. New entrants such as *Lithuania* and *Kazakhstan* suggest a diversification of Ukraine's research partnerships amid the ongoing crisis. Co-funding became more diversified in 2022–2023 (Figure 8). While the EU remained the top source, its share declined from 8.3% to 6.0%. Slovakia (5.3%) rose significantly, overtaking Poland, likely due to increased bilateral support programs. The USA remained a major contributor, while Russia dropped out, having declined to 0.9%. Emerging contributors such as Kazakhstan and Lithuania reflect regional interest in Ukrainian science and demonstrate international solidarity.

Across all three figures, Poland and Slovakia consistently ranked highest or among the top countries, underscoring their geographical closeness, cultural ties, and active support for Ukrainian academics. Germany, China, the UK and the USA also appeared across categories, reflecting strong institutional, research, and funding connections. The disappearance of Russia across all figures in 2022–2023 reflects the severed academic ties due to the war. Overall, there is a noticeable narrowing of foreign co-affiliations toward Central and Western Europe likely driven by displacement dynamics and institutional capacity accompanied by a simultaneous broadening of regional co-authorship and funding partnerships, as evidenced by increased involvement from Lithuania and Kazakhstan.

### 5.3. Foreign co-affiliation, co-authorship and funding in NASU-funded articles

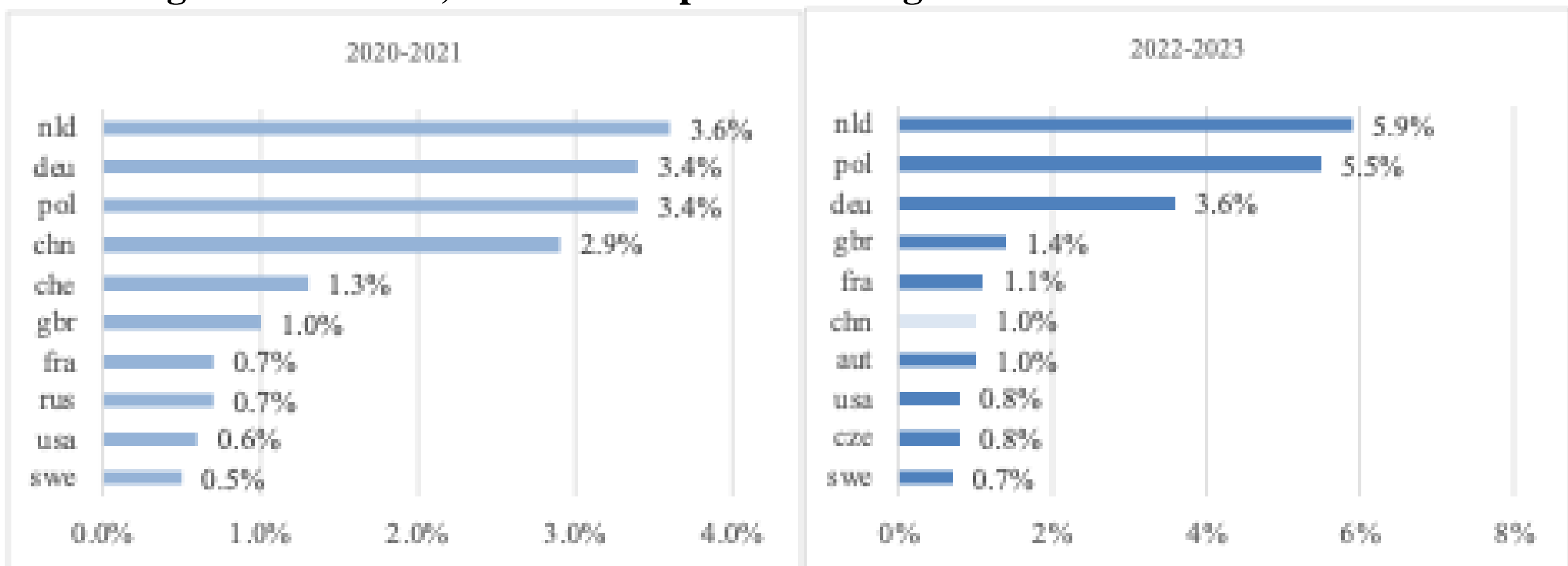

**Fig. 9.** Top ten countries of foreign co-affiliation of Ukrainian scholars in NASU-funded articles

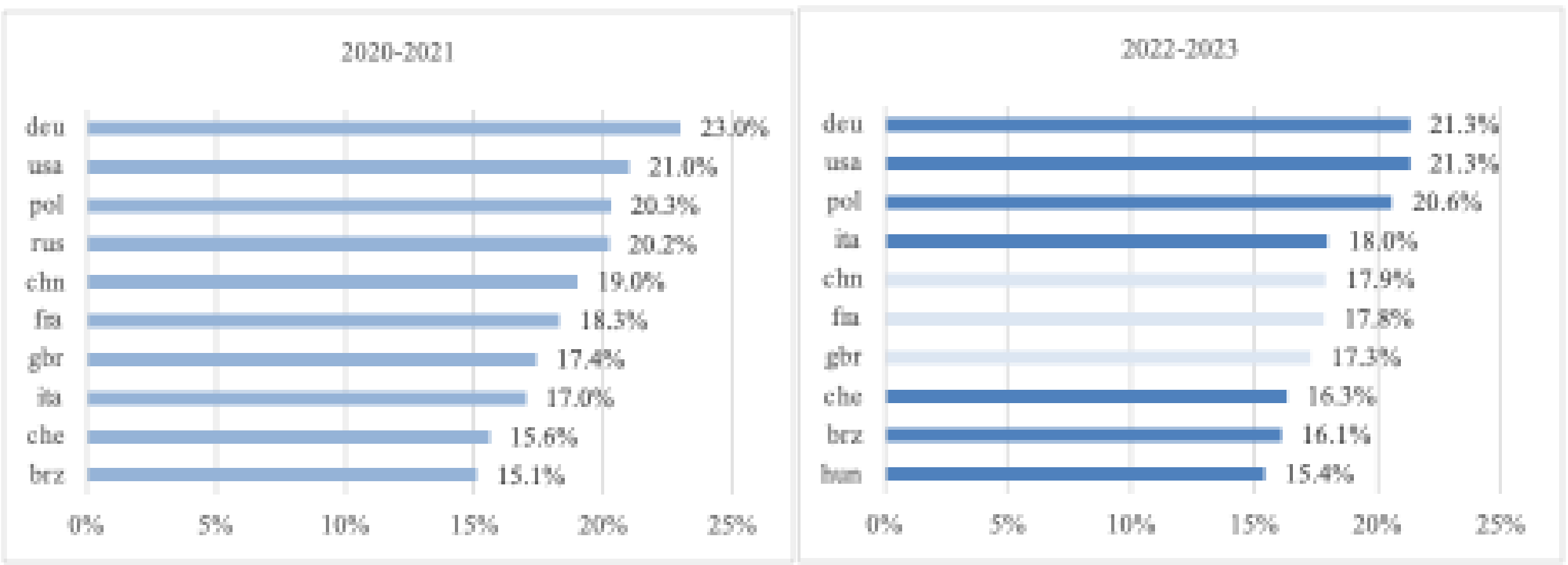

**Fig. 10**. Top ten countries of foreign co-authorship in NASU-funded articles

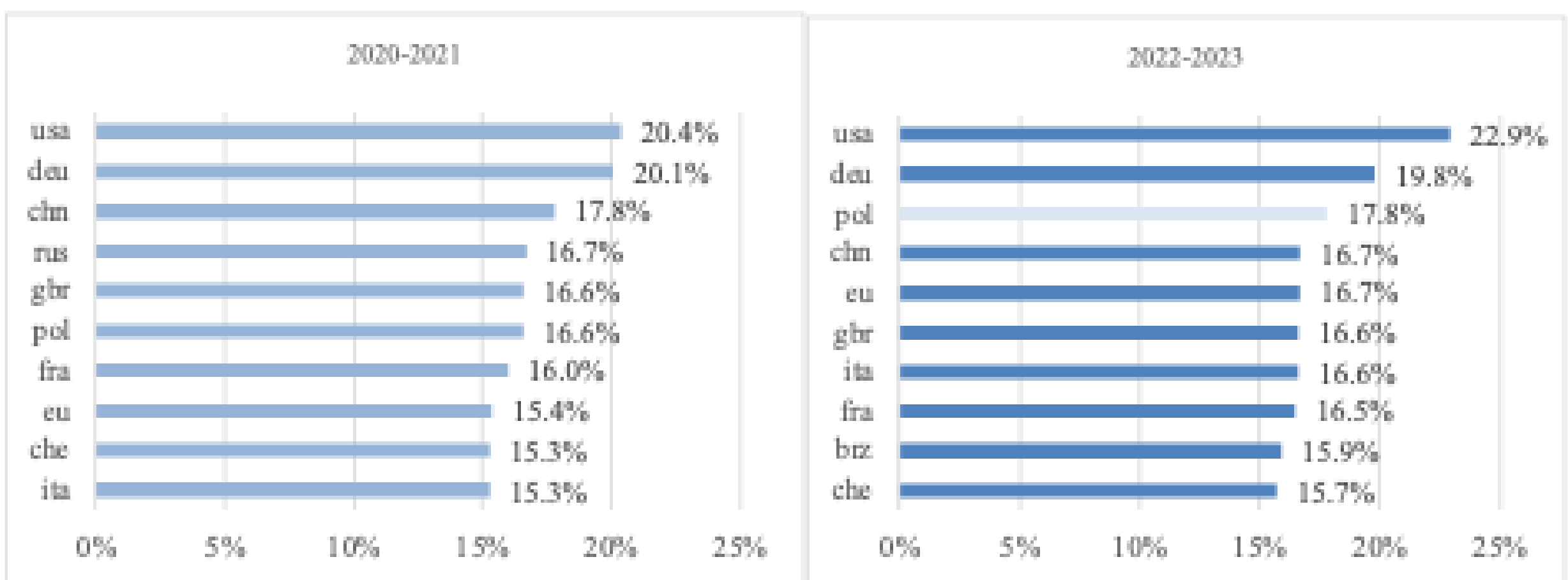

**Fig. 11.** Top ten foreign co-funding countries in NASU-funded articles

Figures 9-11 present trends in foreign co-affiliation, co-authorship, and co-funding in NASU-funded articles across two periods: 2020–2021 and 2022–2023. A comparative analysis reveals both continuities and strategic shifts.

In 2022-2023, the percentage of NASU-funded articles authored by scholars with foreign co-affiliations increased from 18.5% to 22.8%. The Netherlands retained its top position in both periods, increasing its share from 3.6% to 5.9%. Poland and Germany remained important partners, both seeing slight increases. The appearance of Austria and the Czech Republic in 2022–2023 and the disappearance of Russia reflect the shifting geopolitical landscape.

The percentage of NASU-funded articles co-authored with foreign scholars declined from 50.6% (N=637) to 45.0% (N=658). The top three countries Germany, the USA, and Poland retained strong positions across both periods, indicating stable and long-standing research networks. Brazil and China also remained in the top ten, highlighting ongoing partnerships beyond Europe, possibly tied to joint scientific programs or thematic collaboration in STEM fields. While Russia disappeared from the list in 2022–2023, Hungary emerged, indicating some regional diversification. Overall, co-authorship patterns remained broadly consistent, reflecting the resilience of existing networks amid geopolitical disruption.

The USA and Germany remained dominant co-funders of NASU-funded articles, each slightly increasing their shares in 2022–2023. The European Union and China also retained central roles in co-funding, while France, Italy, and Brazil either slightly increased their shares or maintained stable positions. In contrast, Russia dropped out entirely from the list of the top ten. These co-funding patterns indicate a stable and well-supported funding base, increasingly concentrated among key Western partners.

Notably, the average number of foreign co-funding agencies per NASU-funded article declined from 20.9 in 2020–2021 to 18.4 in 2022–2023. Articles that included foreign co-funding typically involved a wide range of co-funding and co-affiliation countries per publication. This trend was largely driven by the field of physics, which dominated these outputs. Given the high costs and technical scale of cutting-edge physics research, such projects often require concentrated resources and large-scale international collaboration frequently centred around major laboratories such as CERN, Fermilab, and SLAC.

Across all three dimensions—co-affiliation, co-authorship, and co-funding—Germany, the USA, and Poland consistently appeared in both periods, confirming their central and sustained roles in close collaboration with NASU. China and Brazil also maintained positions in both co-authorship and co-funding, indicating stable, long-term partnerships. Overall, the figures illustrate a realignment and partial contraction of Ukraine's international academic networks, increasingly focused on reliable Western allies and structured around strategic, high-capacity partnerships.

### 5.4. Foreign co-affiliation, co-authorship and co-funding in NRFU-funded articles

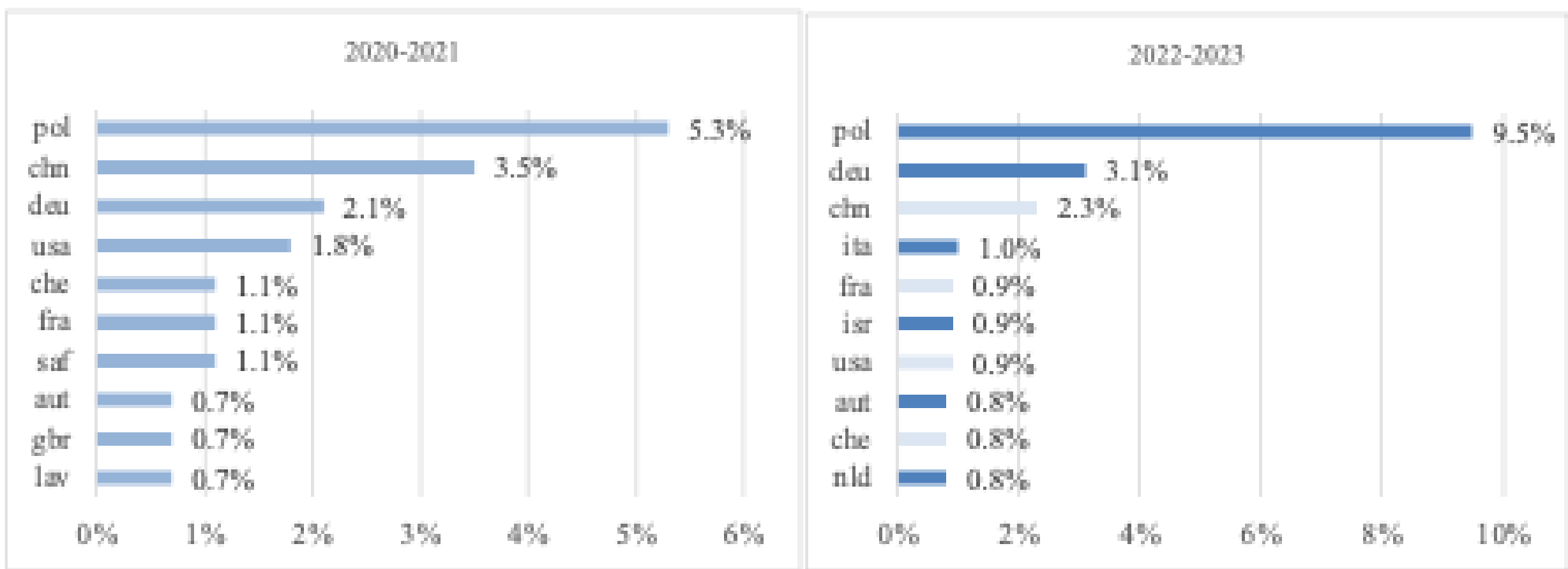

**Fig. 12.** Top ten countries of foreign co-affiliation of Ukrainian scholars in NRFU-funded articles

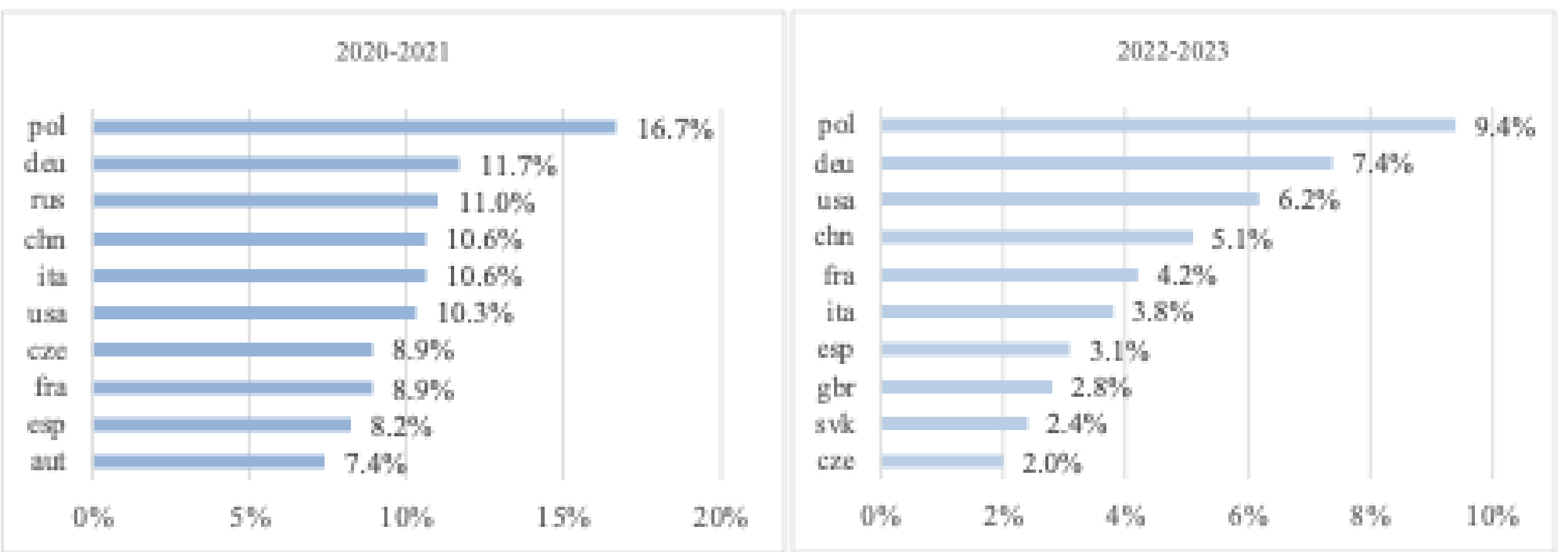

**Fig. 13**. Top ten countries of foreign co-authorship in NRFU-funded articles

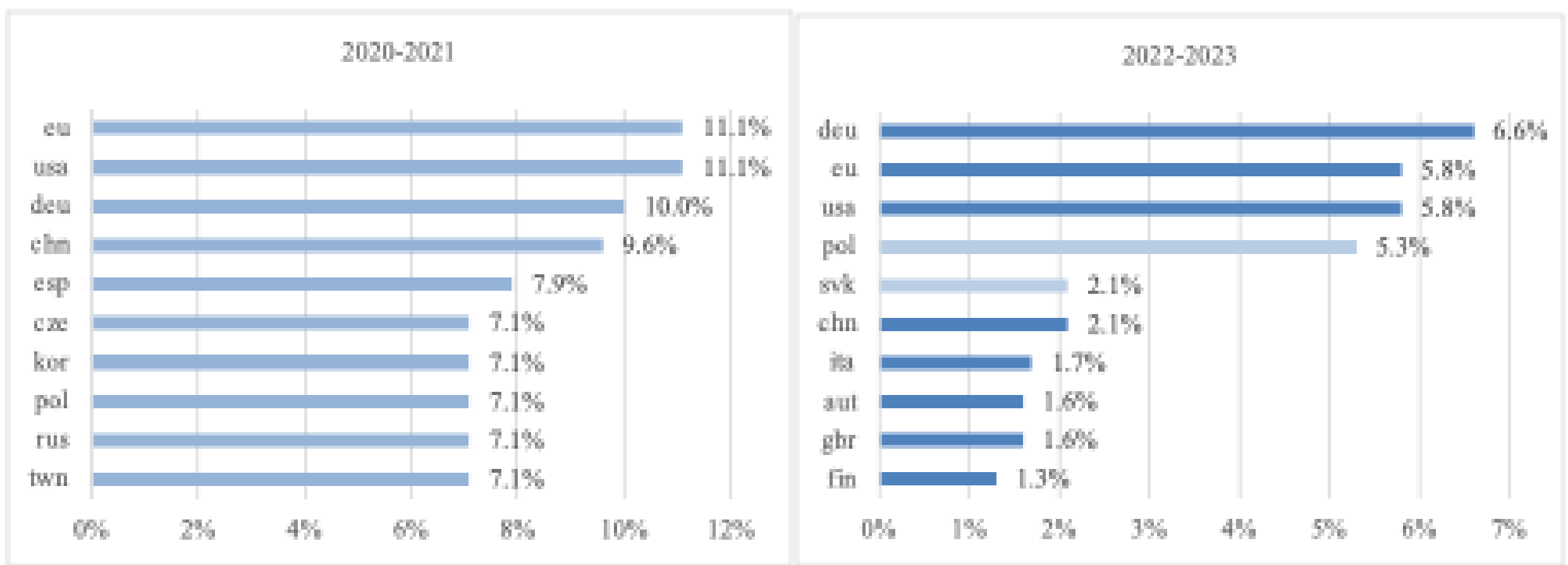

**Fig. 14.** Top ten foreign co-funding countries in NRFU-funded articles

From 2020–2021 to 2022–2023, the percentage of NRFU-funded articles authored by Ukrainian scholars with foreign co-affiliations increased from 17.0% to 22.5%. In contrast, the percentage of articles co-authored with foreign scholars declined from 41.5% to 26.6%.

Figures 12-14 display trends in foreign co-affiliation, foreign co-authorship, and foreign co-funding in NRFU-funded articles from 2020–2021 to 2022–2023. The data reveals substantial shifts in the international dimensions of NRFU-supported research. Between the two periods, Poland nearly doubled, rising from 5.3% to 9.5%, solidifying its position at the top of co-affiliation list. Germany also increased notably (1.8% to 3.1%), while China remained consistently engaged. New entries in 2022–2023 included Israel, Austria, the Netherlands, and France, suggesting a broadening of institutional affiliations. Russia and South Africa, previously in the top ten, were no longer present.

Co-authorship patterns narrowed considerably over the period studied. In 2020–2021, top collaborators included Poland, Spain, Russia, South Korea, Taiwan, and the Czech Republic, reflecting a diverse collaboration landscape. By 2022–2023, Poland, Germany, the USA, and China remained among the leading partners, although their percentage shares declined—likely due to the increase in the total number of NRFU-funded articles. Meanwhile, Russia, South Korea, and Taiwan dropped out of the top ten, while Slovakia and Spain entered the list in 2022–2023.

During the same period, the average number of foreign co-funding agencies per NRFU-funded article dropped sharply from 11.1 in 2020–2021 to just 2.1 in 2022–2023. The EU, the USA, and Germany remained the top co-funders across both periods, although each experienced a decline in share, likely reflecting the overall growth in NRFU-funded articles. Poland emerged as a significant new contributor (5.3%), while Slovakia and Finland appeared for the first time, indicating expanding regional support. In contrast, Russia and Taiwan dropped out of the top co-funding list.

Across all three dimensions—co-affiliation, co-authorship, and co-funding—international collaboration increasingly consolidated around a core group of stable partners, notably Poland, Germany, the USA, and China, reflecting a shift toward politically aligned and well-resourced institutions amid wartime disruptions.

### 5.5. Relationship between funding, authorship patterns and citation impact

Table 3 presents descriptive statistics of field-normalised citation impact (FNCI) for articles published in 2020–2021 and 2022–2023, disaggregated by funding and authorship pattern. Figures 15 and 16 illustrate data from Table 3. Across all funding categories and periods, FNCI distributions are highly skewed, as reflected by large differences between mean and median values and by minimum and maximum FNCI. Mean FNCI values are consistently higher than medians, indicating that a relatively small number of highly cited articles substantially increase average FNCI. This pattern is particularly pronounced for articles with foreign funding (FF) and NASU&FF in both periods as well as MESU in 2022-2023.

**Table 3** Descriptive statistics of FNCI by funding, authorship pattern and period (2020–2021 and 2022–2023)

| Period | Funding | Authorship | Min FNCI | Max FNCI | Mean FNCI | Median FNCI | SD FNCI | Number |
|---|---|---|---|---|---|---|---|---|
| | FF | | **0** | **674** | **1.73** | **0.53** | **13.2** | **3985** |
| | | | **0** | **9.915** | **0.55** | **0.28** | **0.9** | **889** |
| | | FCAF | 0 | 8.32 | 0.78 | 0.37 | 1.46 | 71 |
| | | FCAU | 0 | 9.91 | 0.91 | 0.48 | 1.33 | 235 |
| | MESU | UKR | 0 | 6.98 | 0.42 | 0.23 | 0.62 | 617 |
| | | | **0** | **13.4** | **0.77** | **0.43** | **1.2** | **466** |
| | | FCAF | 0 | 13.45 | 0.90 | 0.60 | 1.35 | 157 |
| | | FCAU | 0 | 13.45 | 0.81 | 0.48 | 1.19 | 392 |
| | MESU&FF | UKR | 0 | 7.11 | 0.48 | 0.21 | 1.00 | 58 |
| | | | **0** | **6.12** | **0.36** | **0.18** | **0.5** | **728** |
| | | FCAF | 0 | 3.92 | 0.59 | 0.45 | 0.67 | 60 |
| | | FCAU | 0 | 6.12 | 0.63 | 0.48 | 0.78 | 161 |
| | NASU | UKR | 0 | 3.67 | 0.28 | 0.17 | 0.40 | 545 |
| | | | **0** | **14.3** | **1.28** | **0.63** | **1.9** | **532** |
| | | FCAF | 0 | 13.43 | 1.40 | 0.62 | 2.27 | 173 |
| | | FCAU | 0 | 14.31 | 1.40 | 0.76 | 1.97 | 475 |
| | NASU&FF | UKR | 0 | 1.71 | 0.29 | 0.19 | 0.36 | 47 |
| | | | **0** | **8.32** | **0.63** | **0.31** | **0.9** | **195** |
| | | FCAF | 0 | 8.32 | 0.76 | 0.32 | 1.72 | 22 |
| | | FCAU | 0 | 8.32 | 0.79 | 0.35 | 1.21 | 90 |
| | NRFU | UKR | 0 | 4.28 | 0.50 | 0.28 | 0.72 | 104 |
| | | | **0** | **5.3** | **1.05** | **0.83** | **0.9** | **87** |
| | | FCAF | 0 | 3.92 | 1.11 | 0.92 | 0.98 | 26 |
| | | FCAU | 0 | 5.30 | 1.14 | 0.91 | 0.98 | 74 |
| | NRFU&FF | UKR | 0 | 1.22 | 0.54 | 0.37 | 0.45 | 13 |
| | | | 0 | **137** | **0.28** | **0.09** | **1.1** | **24353** |
| | | FCAF | 0 | 23.31 | 0.57 | 0.24 | 1.34 | 780 |
| 2020-2021 | | FCAU | 0 | 137.23 | 0.65 | 0.26 | 2.53 | 4041 |
| | WF | UKR | 0 | 11.06 | 0.20 | 0.08 | 0.42 | 20047 |
| | FF | | **0** | **348** | **1.55** | **0.51** | **7.6** | **4247** |
| | | | | **32.1** | **0.93** | **0.3** | **2.3** | **1004** |
| | | FCAF | 0 | 32.07 | 2.68 | 0.70 | 5.07 | 145 |
| | | FCAU | 0 | 32.07 | 1.45 | 0.51 | 3.01 | 264 |
| | MESU | UKR | 0 | 11.52 | 0.54 | 0.26 | 0.94 | 676 |
| | | | **0** | **17.2** | **0.84** | **0.42** | **1.6** | **498** |
| | | FCAF | 0 | 15.04 | 0.93 | 0.47 | 1.67 | 212 |
| | | FCAU | 0 | 17.21 | 0.94 | 0.51 | 1.67 | 408 |
| | MESU&FF | UKR | 0 | 2.72 | 0.35 | 0.22 | 0.53 | 59 |
| | | | | **2.65** | **0.33** | **0.17** | **0.4** | **871** |
| | | FCAF | | 2.32 | 0.53 | 0.40 | 0.57 | 77 |
| | | FCAU | | 2.51 | 0.49 | 0.34 | 0.52 | 164 |
| | NASU | UKR | | 2.65 | 0.28 | 0.14 | 0.39 | 667 |
| | | | | **21.8** | **1.19** | **0.56** | **2.2** | **591** |
| | | FCAF | 0 | 21.80 | 1.34 | 0.43 | 2.90 | 256 |
| | | FCAU | 0 | 21.80 | 1.35 | 0.70 | 2.33 | 494 |
| | NASU&FF | UKR | 0 | 1.29 | 0.36 | 0.28 | 0.29 | 48 |
| | | | | **11.0** | **0.61** | **0.3** | **1.1** | **627** |
| | | FCAF | 0 | 10.95 | 0.86 | 0.42 | 1.49 | 90 |
| | | FCAU | 0 | 10.95 | 1.04 | 0.49 | 1.66 | 192 |
| | NRFU | UKR | 0 | 4.33 | 0.42 | 0.26 | 0.55 | 434 |
| | | | | **7.64** | **0.77** | **0.51** | **1** | **288** |
| | | FCAF | 0 | 7.64 | 0.89 | 0.56 | 1.18 | 116 |
| | | FCAU | 0 | 7.64 | 0.81 | 0.56 | 0.99 | 251 |
| | NRFU&FF | UKR | 0 | 4.69 | 0.50 | 0.14 | 0.97 | 32 |
| | | | | **462** | **0.39** | **0.14** | **3.3** | **21825** |
| | | FCAF | 0 | 16.23 | 0.74 | 0.32 | 1.27 | 1039 |
| | | FCAU | 0 | 461.54 | 0.95 | 0.34 | 7.67 | 3855 |
| 2022-2023 | WF | UKR | 0 | 11.60 | 0.26 | 0.09 | 0.62 | 17533 |

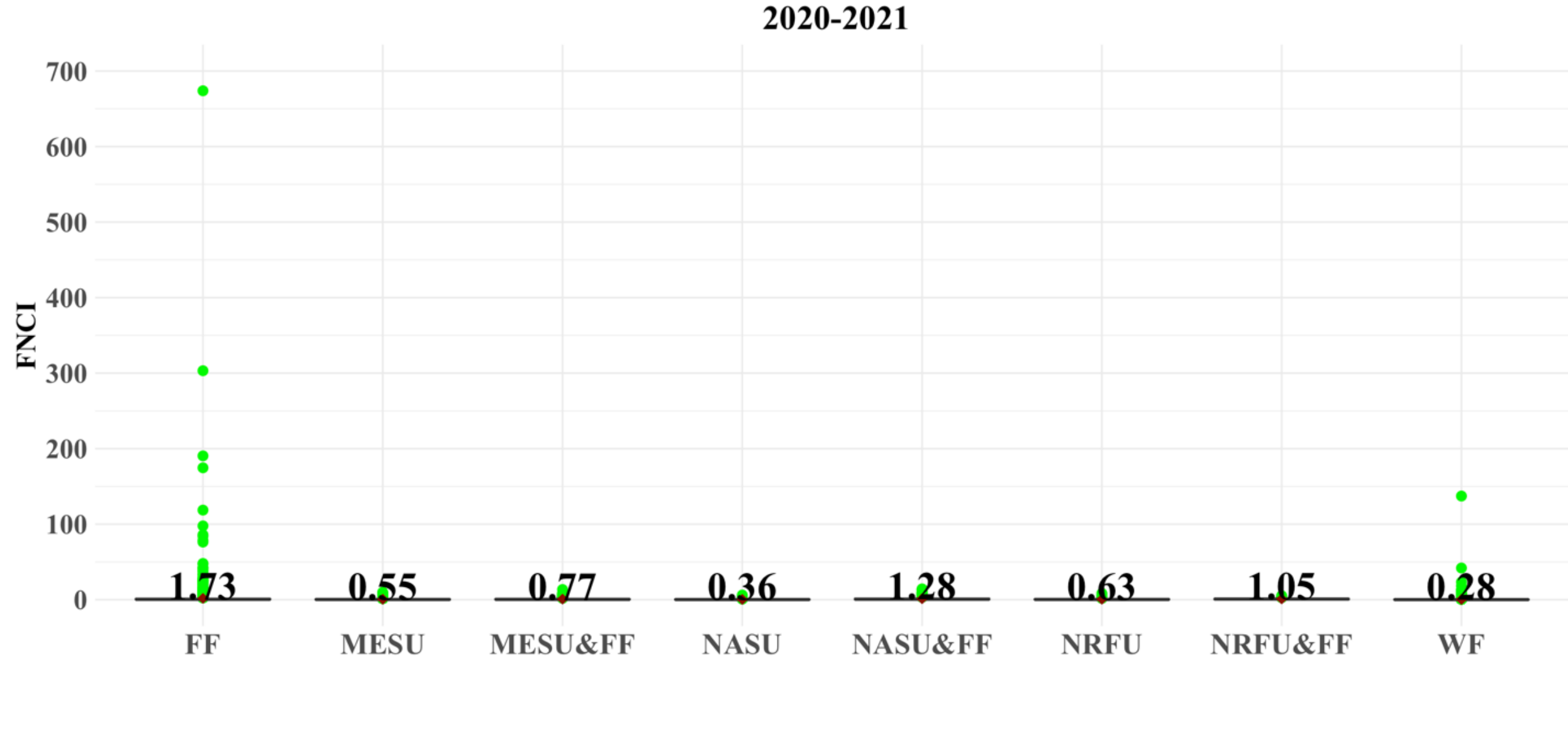

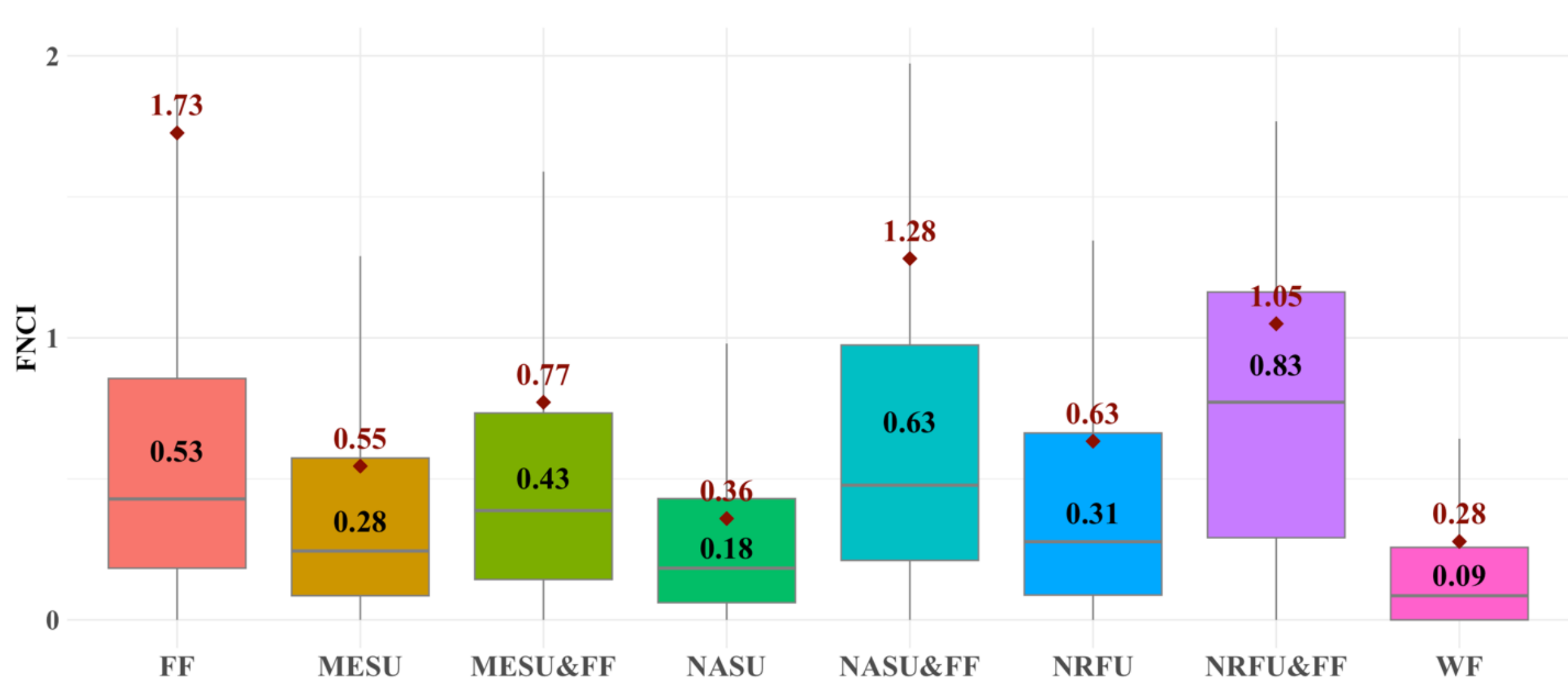

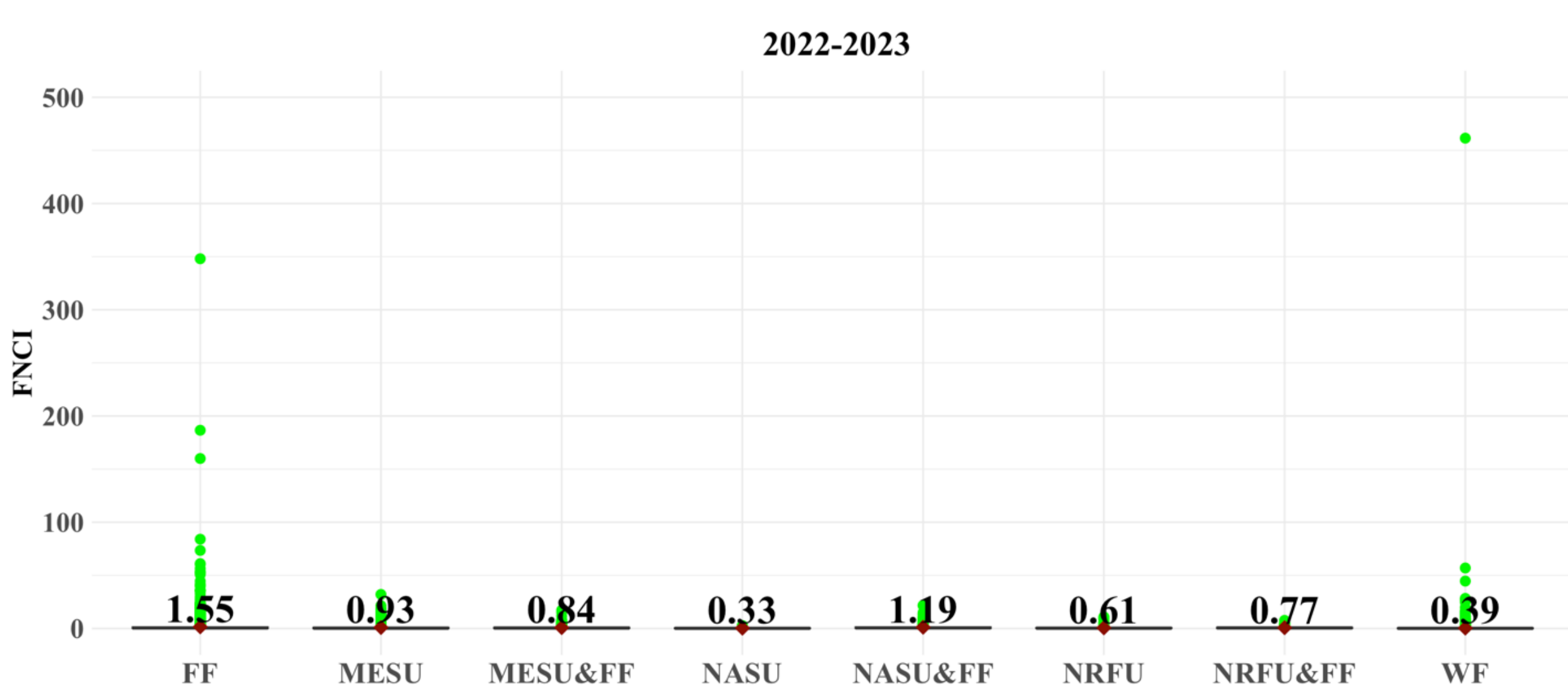

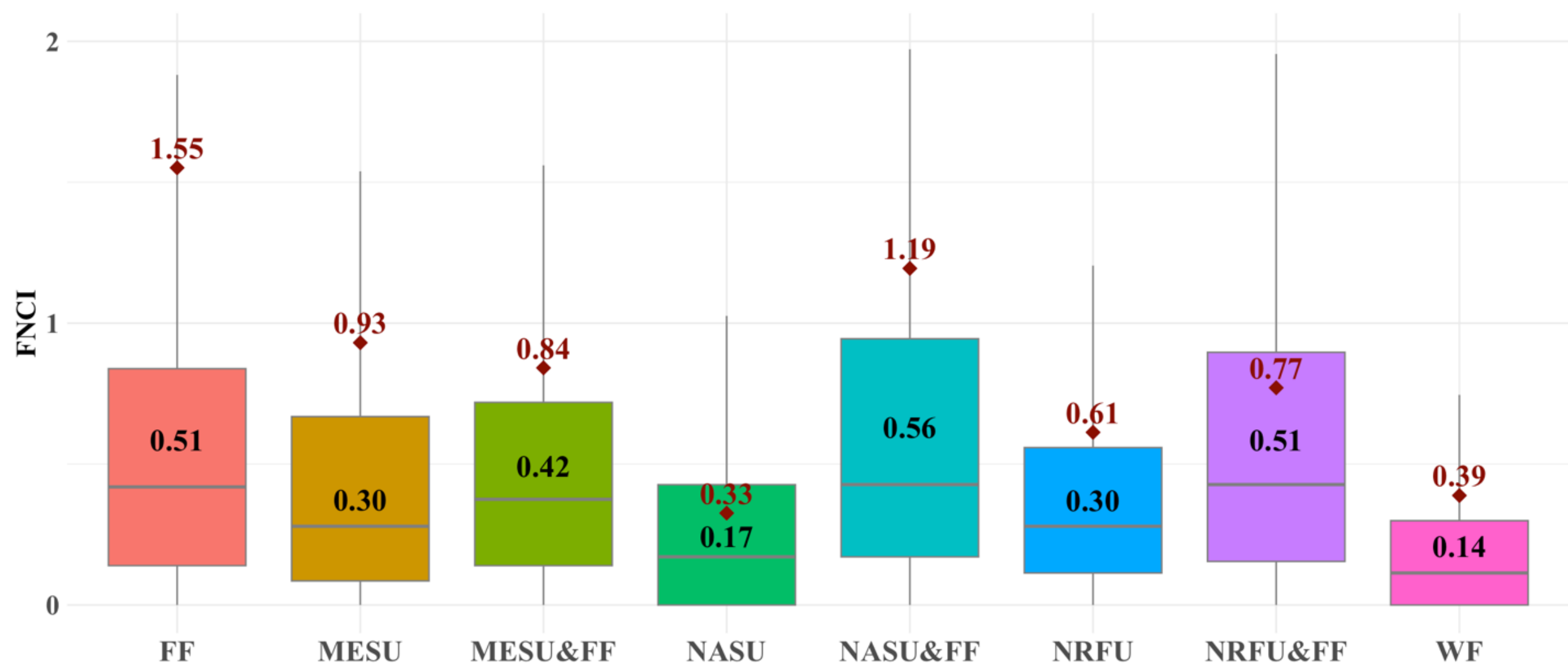

**Fig. 15.** FNCI across funding patterns (MESU, NASU and NRFU refer to funding provided by Ukrainian agencies without foreign co-funding)

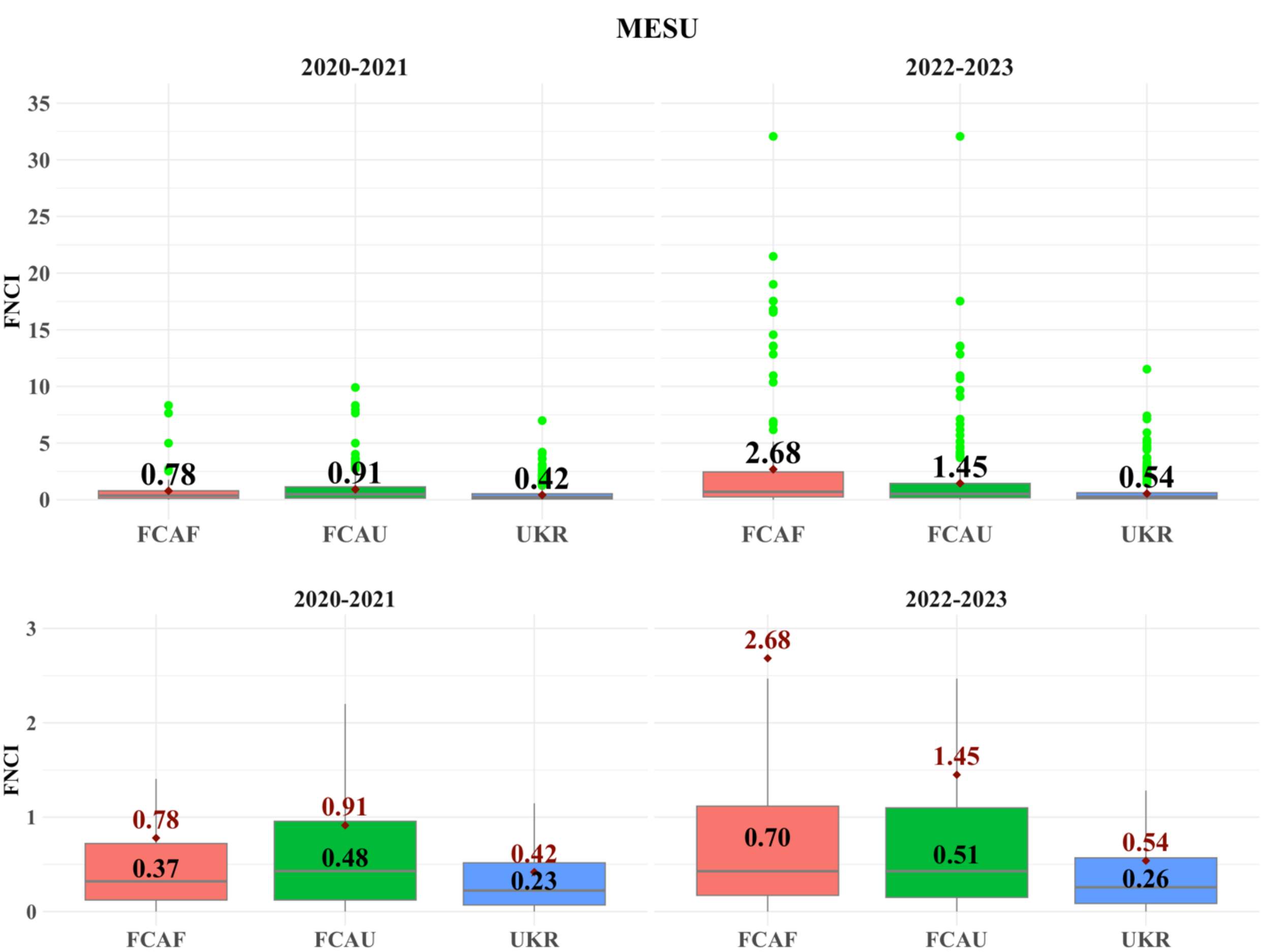

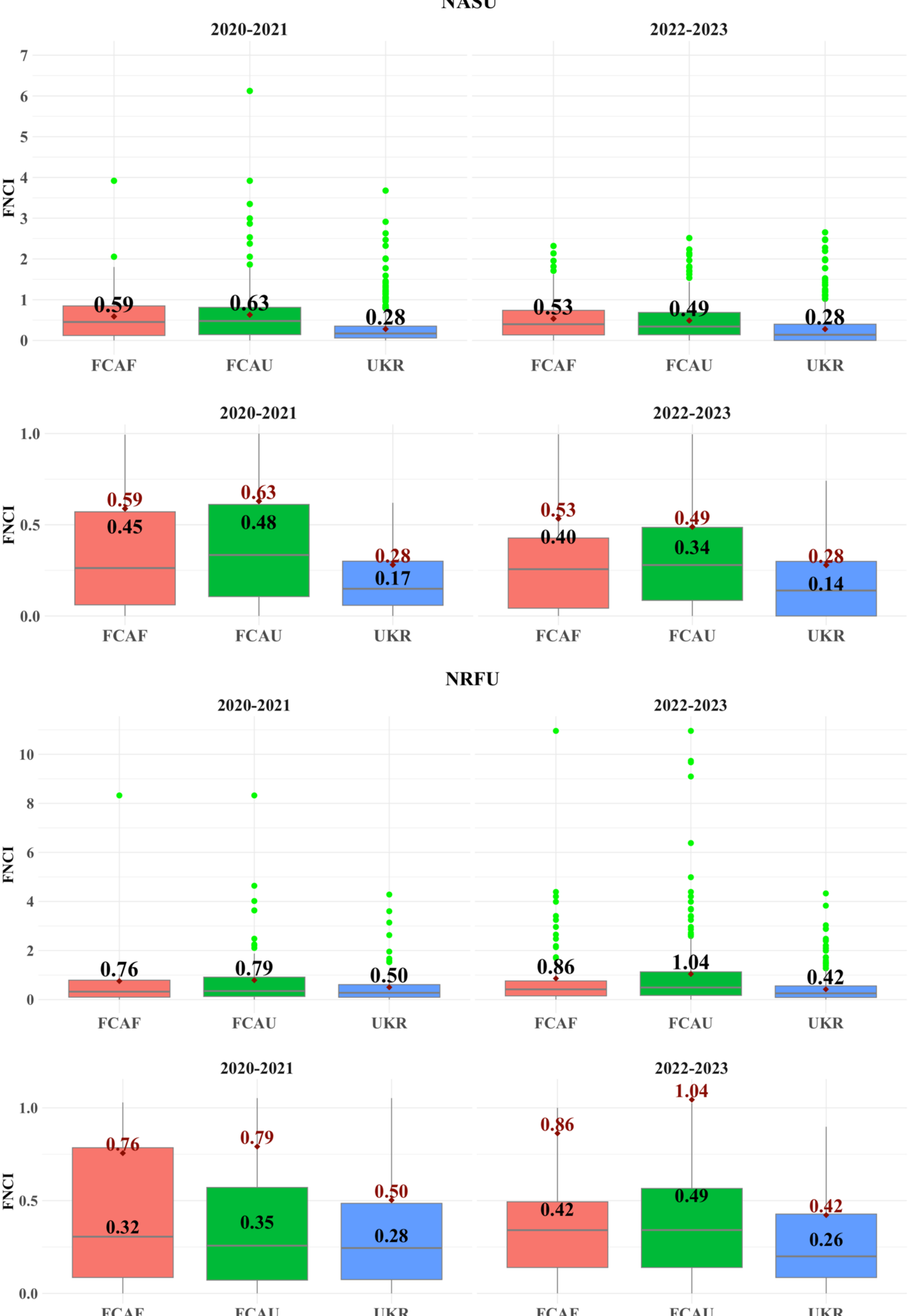

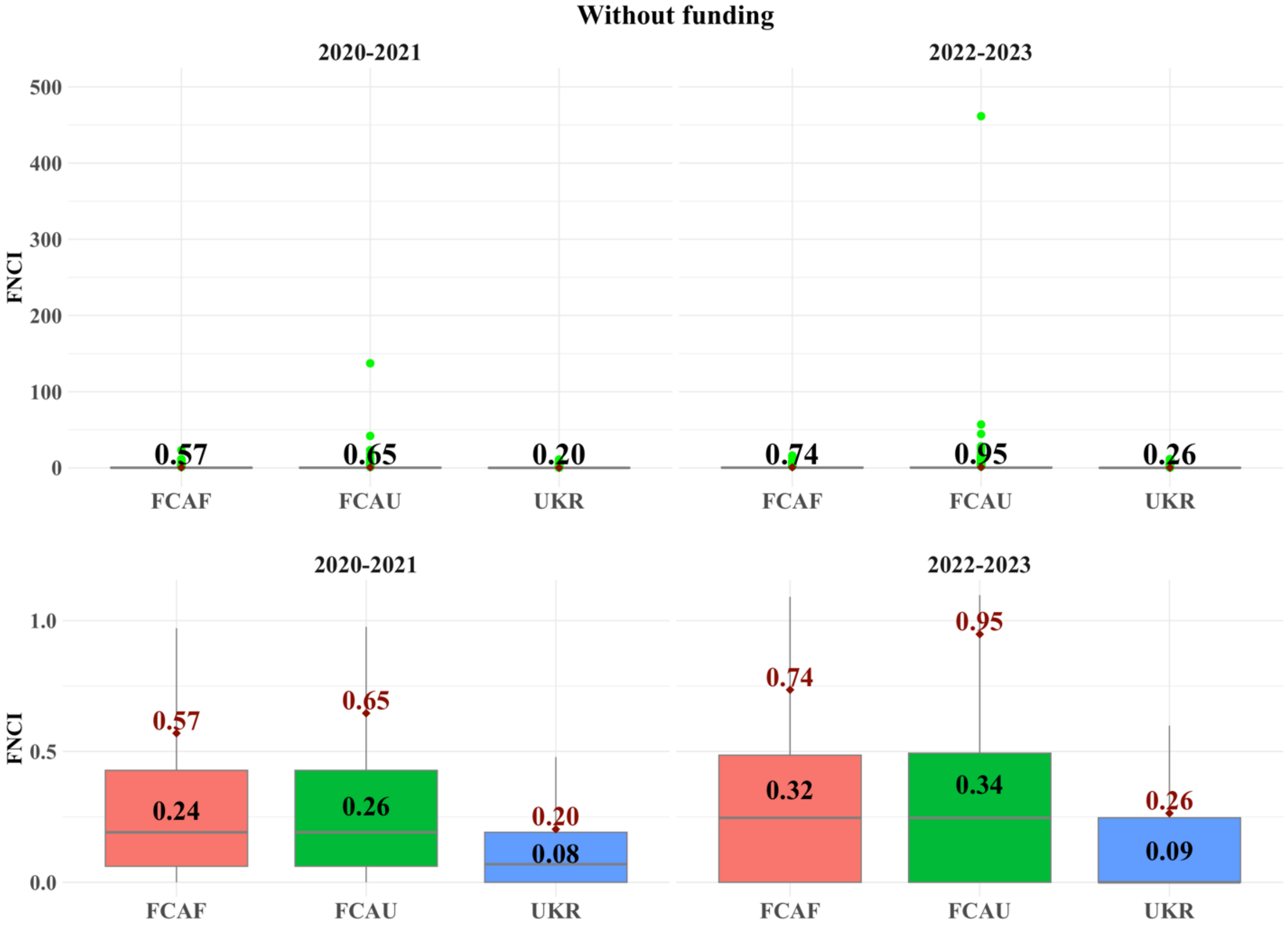

Fig. 16. Field-Normalised Citation Impact (FNCI) across different types of funding and authorship.

In both periods, articles funded solely by foreign agencies (FF) had the highest average FNCI citation impact (1.73 and 1.55 correspondingly). They also featured the greatest number of highly cited articles (Figure 15) and the highest maximum FNCI. They were followed by articles with joint Ukrainian and foreign funding in 2020-2021. However, in 2022-2023, the average FNCI of MESU-funded articles exceeded the average FNCI of MESU&FF, having almost doubled from 0.55 to 0.93. Table 3 and Fig 16 show that it occurred because of the rise in the average FNCI of articles authored by scholars with foreign co-affiliations and internationally co-authored articles. The average of FNCI of MESU-funded articles authored by scholars with foreign co-affiliation rose dramatically from 0.78 in 2020–2021 to 2.68 in 2022–2023. Internationally co-authored MESU-funded articles also saw a moderate rise, with average FNCI increasing from 0.91 to 1.45 over the same period. By contrast, articles authored solely by scholars with Ukrainian affiliations continued to exhibit significantly lower citation impact. The benefit of foreign co-funding also diminished for average FNCI of NFRU&FF articles in 2022–2023 from 1.05 to 0.77, narrowing the gap between NRFU (0.63 and 0.61) and NRFU&FF. Overall, foreign co-funding consistently produced the highest citation impact, except for MESU in 2022-2023, while Ukrainian-only funding schemes (especially NASU) lagged behind. The increase in MESU's performance in 2022–2023 stands out as a notable exception.

NASU-funded articles consistently had the lowest average FNCI among articles funded by three agencies in both periods (0.36 and 0.33). In 2022, the average FNCI of articles without funding (0.28 and 0.39) marginally exceeded the average FNCI of NASU-funded articles.

Table 3 and Fig 16 show that articles involving foreign co-affiliation (FCAF) and foreign co-authorship (FCAU) generally showed higher average FNCI than domestically authored (UKR) articles across funding agencies and periods. Median FNCI values remain below 1 in most categories, indicating that the typical article performs at or below the global citation average.

Notably, the average FNCI of unfunded articles authored by scholars with Ukrainian affiliations closely mirrored those of NASU-funded articles—0.20 vs. 0.28 in 2020–2021 and 0.26 vs. 0.28 in 2022–2023. While in 2020-2021, the average FNCI of articles without funding (0.57; 0.65) and NASU-funded articles (0.59; 0.63) authored by scholars with foreign co-affiliations or international co-authorship was almost equal, in 2022-2023, average FNCI of articles without funding exceeded those of NASU-funded articles (0.74 vs 0.53 for articles with foreign co-affiliations and 0.95 vs 0.45 for internationally co-authored articles), underscoring the strong influence of international collaboration on research visibility regardless of funding status.

These findings affirm that international engagement, more than domestic funding alone, plays a critical role in enhancing the global impact of Ukrainian research.

Given the overdispersion of FNCI data, a negative binomial regression model with interactions between period (2020–2021 vs. 2022–2023), funding, and authorship pattern was used to explore the relationship between funding, authorship patterns, and FNCI across two periods. Table 4 summarises the results of the statistical test. The intercept represents NASU-funded articles in 2020–2021 with Ukrainian authorship, with an expected FNCI of 0.28 (IRR = 0.28), indicating a relatively low baseline citation count.

Funding type overall showed strong effects: foreign funding (FF) was strongly associated with higher FNCI ($\beta = 1.718$, $p < 0.001$), followed by MESU ($\beta = 0.398$, $p < 0.001$) and NRFU ($\beta = 0.582$, $p = 0.002$). Without funding (WF) in 2020–2021 was negatively associated with FNCI ($\beta = -0.325$, $p < 0.001$), but the positive period × WF interaction in 2022–2023 partially offsets this effect. MESU and NRFU funding (without foreign co-funding) significantly increase citation impact (IRR ≈ 1.49, $p < 0.001$; IRR ≈ 1.79, $p = 0.002$, respectively) compared to NASU funding. These results suggest that certain national funding agencies (MESU, NRFU) are associated with higher impact outcomes, independent of authorship patterns.

Articles published in 2022–2023 with NASU funding and Ukrainian authorship showed no significant change compared to the baseline ($\beta = -0.006$, $p = 0.959$). In contrast, articles without funding (WF) in 2022–2023 with Ukrainian authorship exhibited a positive interaction effect ($\beta = 0.270$, $p = 0.031$), suggesting a modest increase in FNCI compared to NASU-funded articles (IRR = 1.31, $p = 0.031$). However, the combined effects indicate that unfunded articles still remain slightly below NASU-funded articles in terms of citation impact, suggesting a partial convergence rather than a reversal.

NASU FCAF (IRR ≈ 2.10, $p = 0.001$) and FCAU (IRR ≈ 2.24, $p < 0.001$) articles show higher citation rates compared to UKR, yet the relative gains over WF FCAF (IRR ≈ 1.34) and WF FCAU (IRR ≈ 1.42) are not statistically robust. Significant gains in FNCI are shown for the 2022-2023× MESU × FCAF authorship interaction ($\beta = 1.075$, $p = 0.005$).

When compulsory Ukrainian authorship is taken into account, the effects of foreign co-funding differ across funding agencies. MESU-funded articles co-funded with foreign agencies (MESU&FF) show a clear and statistically significant citation advantage compared to MESU-funded articles without foreign co-funding (IRR ≈ 1.72, $p = 0.027$), and this effect does not vary significantly across authorship configurations. In contrast, NRFU-funded articles with foreign co-funding (NRFU&FF) display a positive but statistically non-robust effect (IRR ≈ 1.92, $p = 0.166$). For NASU, foreign co-funding alone does not increase citation impact (IRR ≈ 1.03, $p = 0.92$); however, when combined with international authorship ties, citation impact rises significantly (FCAF: IRR ≈ 2.30, $p = 0.034$; FCAU: IRR ≈ 2.15, $p = 0.027$). The results suggest that international collaboration patterns, rather than NASU funding alone, drive citation impact. The effectiveness of foreign co-funding depends on both its presence and how it interacts with authorship structures.

Overall, McFadden pseudo-$R^2 = 0.11$, reflecting moderate explanatory power, and predicted citation counts highlight substantial heterogeneity across funding, authorship, and period. Given the modest explanatory power of the model, these results should be interpreted as average group-level tendencies rather than deterministic effects.

**Table 4.** Negative binomial regression: FNCI, funding and authorship patterns

| Coefficient | Estimate | Std_Error | p_value | IRR | IRR_lower | IRR_upper | Significant |
|---|---|---|---|---|---|---|---|
| (Intercept) | -1,27 | 0,09 | <0.001 | 0,28 | 13,20 | 0,09 | *** |
| 2022-2023 | -0,01 | 0,12 | 0.959 | 0,99 | 1,14 | 1,12 | |
| FF | 1,72 | 0,09 | <0.001 | 5,57 | 0,04 | 31,78 | *** |
| MESU | 0,40 | 0,12 | <0.001 | 1,49 | 0,51 | 2,45 | *** |
| MESU&FF | 0,54 | 0,24 | 0.027 | 1,72 | 0,44 | 3,69 | * |
| NASU&FF | 0,03 | 0,32 | 0.920 | 1,03 | 1,29 | 1,46 | |
| NRFU | 0,58 | 0,19 | 0.002 | 1,79 | 0,39 | 3,79 | ** |
| NRFU&FF | 0,65 | 0,47 | 0.166 | 1,92 | 0,45 | 5,76 | |
| WF | -0,33 | 0,09 | <0.001 | 0,72 | 2,07 | 0,58 | *** |
| FCAF | 0,74 | 0,23 | 0.001 | 2,10 | 0,29 | 5,36 | ** |
| FCAU | 0,81 | 0,15 | <0.001 | 2,24 | 0,24 | 5,68 | *** |
| 2022-2023:FF | -0,03 | 0,12 | 0.836 | 0,97 | 1,19 | 1,08 | |
| 2022-2023:MESU | 0,26 | 0,16 | 0.098 | 1,30 | 0,70 | 1,94 | |
| 2022-2023:MESU&FF | -0,33 | 0,36 | 0.365 | 0,72 | 2,73 | 0,76 | |
| 2022-2023:NASU&FF | 0,22 | 0,43 | 0.618 | 1,24 | 1,01 | 2,35 | |
| 2022-2023:NRFU | -0,17 | 0,23 | 0.455 | 0,85 | 1,74 | 0,90 | |
| 2022-2023:NRFU&FF | -0,07 | 0,57 | 0.897 | 0,93 | 2,03 | 1,53 | |
| 2022-2023:WF | 0,27 | 0,12 | 0.031 | 1,31 | 0,67 | 1,92 | * |
| 2022-2023:FCAF | -0,09 | 0,31 | 0.766 | 0,91 | 1,63 | 1,14 | |
| 2022-2023:FCAU | -0,25 | 0,22 | 0.264 | 0,78 | 2,02 | 0,77 | |
| MESU:FCAF | -0,12 | 0,30 | 0.693 | 0,89 | 1,69 | 1,07 | |
| MESU&FF:FCAF | -0,11 | 0,34 | 0.741 | 0,89 | 1,76 | 1,13 | |
| NASU&FF:FCAF | 0,83 | 0,39 | 0.034 | 2,30 | 0,29 | 7,60 | |
| NRFU:FCAF | -0,33 | 0,43 | 0.437 | 0,72 | 2,94 | 0,80 | |
| NRFU&FF:FCAF | -0,02 | 0,58 | 0.978 | 0,98 | 1,84 | 1,73 | |
| WF:FCAF | 0,29 | 0,24 | 0.216 | 1,34 | 0,71 | 2,25 | |
| MESU:FCAU | -0,03 | 0,19 | 0.893 | 0,97 | 1,28 | 1,15 | |
| MESU&FF:FCAU | -0,29 | 0,28 | 0.304 | 0,75 | 2,36 | 0,75 | |
| NASU&FF:FCAU | 0,77 | 0,35 | 0.027 | 2,15 | 0,32 | 6,35 | * |
| NRFU:FCAU | -0,35 | 0,28 | 0.201 | 0,70 | 2,63 | 0,66 | |
| NRFU&FF:FCAU | -0,06 | 0,51 | 0.908 | 0,94 | 1,87 | 1,49 | |
| WF:FCAU | 0,35 | 0,16 | 0.026 | 1,42 | 0,59 | 2,33 | * |
| 2022-2023:MESU:FCAF | 1,07 | 0,38 | 0.005 | 2,93 | 0,18 | 12,03 | ** |
| 2022-2023:MESU&FF:FCAF | 0,46 | 0,48 | 0.342 | 1,58 | 0,66 | 3,99 | |
| 2022-2023:NASU&FF:FCAF | -0,16 | 0,53 | 0.761 | 0,85 | 2,33 | 1,24 | |
| 2022-2023:NRFU:FCAF | 0,40 | 0,51 | 0.431 | 1,49 | 0,76 | 3,62 | |
| 2022-2023:NRFU&FF:FCAF | -0,06 | 0,70 | 0.936 | 0,95 | 2,25 | 1,80 | |
| 2022-2023:WF:FCAF | 0,08 | 0,32 | 0.792 | 1,09 | 1,17 | 1,62 | |
| 2022-2023:MESU:FCAU | 0,46 | 0,27 | 0.092 | 1,58 | 0,54 | 3,20 | |
| 2022-2023:MESU&FF:FCAU | 0,73 | 0,42 | 0.081 | 2,08 | 0,36 | 6,38 | |
| 2022-2023:NASU&FF:FCAU | 0,01 | 0,48 | 0.990 | 1,01 | 1,59 | 1,63 | |
| 2022-2023:NRFU:FCAU | 0,70 | 0,34 | 0.042 | 2,01 | 0,36 | 5,55 | * |
| 2022-2023:NRFU&FF:FCAU | -0,02 | 0,62 | 0.977 | 0,98 | 1,93 | 1,80 | |
| 2022-2023:WF:FCAU | 0,37 | 0,22 | 0.102 | 1,44 | 0,61 | 2,57 | |

## 6. Conclusions

This study explored funding, authorship patterns and citation impact of articles funded by the Ministry of Education and Science of Ukraine (MESU), the National Academy of Sciences of Ukraine (NASU) and the National Research Foundation of Ukraine (NRFU) and published two years before (2020-2021) and two years during (2022-2023) the Russia's full scale invasion of Ukraine.

Overall, the study findings revealed that the proportion of Ukrainian articles acknowledging research funding has increased between two periods from, with a notable rise in the share of articles funded by MESU, NASU, and NRFU from 8.6% to 11.9%. The results show that above 30% of NRFU and MESU funding articles also were supported by foreign co-funding, with this share reaching 40.4% for NASU where prevail physical sciences.

In both periods, foreign co-funding largely came from the EU, USA, Germany, China, and Poland. The presence of EU, USA, Germany and China among most frequently acknowledged co-funding countries aligns with prior research indicating the significant role of these countries in global scientific collaboration, driven by their high levels of R&D investment (Statista, 2024; Sargent, 2022; Miao et al., 2023). Poland's growing importance, especially for MESU- and NRFU-funded articles, reflects long-term bilateral collaboration intensified since Russia's invasion of Ukraine in 2014 (OECD, 2022; Kiselyova & Ivashchenko, 2024). Meanwhile, the emergence of Slovakia, Lithuania, and Kazakhstan as significant co-funders in 2022–2023 indicates strengthened scientific collaboration. In 2020-2021, Russia was among the top ten foreign co-funding countries for articles funded by three Ukrainian agencies, but not in 2022-2023. A limitation of this study is that we did not investigate whether this funding reflected bilateral arrangements between Ukraine and Russia or multilateral collaborations typical of the physical sciences, which are dominant at NASU

The share of internationally co-authored articles was higher among those with foreign co-funding revealing that research funding stengthens international collaboration. Around half of the articles funded by each agency was co-authored with foreign scholars or involved Ukrainian researchers with foreign co-affiliations. The share of authors with foreign affiliations increased notably in 2022–2023. Arguably this can be attributed to the effects of the war, which entailed Ukrainian scholoars moving abroad and the total number of articles authored by Ukrainian scholars increased (Hladchenko, 2025). Poland maintained its leading position among countries of foreign co-affiliation in MESU- and NRFU-funded articles, while the Netherlands was the top country of foreign co-affiliation for NASU-funded articles. This occurs because the Netherlands has actively recruited scholars in the physical sciences from abroad due to a shortage of physicists (van Hattem, 2011).However, the overall rate of international co-authorship declined slightly in articles funded by three Ukrainian agencies. Despite this, Poland, Germany, and the USA remained central collaborators.

Overall, funded articles exhibited higher average field-normalised citation impact (FNCI) than unfunded ones, in line with previous studies (Wang & Shapira, 2015; Trochim et al., 2008; Sandström, 2009; Campbell et al., 2010; Roshani et al., 2021; Tian et al., 2025). However, this advantage was not uniform and depended strongly on both the funding source and authorship configuration. Foreign co-funding and international co-authorship and co-affiliations were consistently associated with higher citation impact. In particular, foreign co-affiliations had a pronounced positive effect for MESU-funded articles in 2022–2023, where their FNCI exceeded that of articles jointly funded by MESU and foreign agencies. This pattern supports earlier findings that foreign co-affiliations facilitate access to complementary resources and enhance research visibility (Lander, 2015). At the same time, the presence of foreign co-affiliations complicates institutional and national performance assessment, as attribution of research outcomes becomes increasingly diffuse (Hottenrott & Lawson, 2017; Safaei et al., 2016; Gingras, 2014).

NASU funding is associated with only modest differences in citation impact relative to unfunded articles. These effects are small and not consistently significant across authorship patterns, and they weaken in relative terms in 2022–2023 as unfunded articles partially converge in citation performance. Taken together, these findings suggest limited citation-related efficiency of NASU funding when evaluated through field-normalised citation impact. While the results should be interpreted as average group-level tendencies rather than deterministic effects, they raise important questions about the effectiveness of current funding allocation mechanisms and evaluation criteria, highlighting the need for evidence-based reform of Ukraine's research funding system, which is crutial in the context of ongoing war.

A limitation of this study is that we did not explore differences across disciplines, which may influence the effects of funding, authorship patterns, and citation impact of articles funded by three funding agencies.

Overall, the findings underscore the continuing importance of jointly funded projects and international collaboration, suggesting that joint funding initiatives with foreign agencies should become a strategic priority in Ukrainian science policy. Recalibrating domestic funding mechanisms to better support international engagement may not only enhance research visibility and citation impact but also strengthen the resilience of Ukraine's scientific system under wartime conditions.

Finally, the methodological challenges encountered in this study—particularly in identifying articles funded by Ukrainian agencies—underscore the need for a centralised national funding database. A transparent and comprehensive repository documenting grants from MESU, NASU, and NRFU, along with information on funded researchers and research outputs, would significantly enhance the tracking, evaluation, and coordination of research efforts in Ukraine.